\documentclass[lettersize,journal]{IEEEtran}
\usepackage{amsmath,amsfonts}
\usepackage[linesnumbered,algoruled,boxed,lined]{algorithm2e}

\usepackage{array}
\usepackage[caption=false,font=normalsiz
e,labelfont=sf,textfont=sf]{subfig}
\usepackage{textcomp}
\usepackage{stfloats}
\usepackage{url}
\usepackage{verbatim}
\usepackage{graphicx}
\usepackage{cite}
\usepackage{amssymb}   
\usepackage{enumitem}
\usepackage{xcolor}
\usepackage{amsmath}
\usepackage{tabularx}
\usepackage{makecell} 
\usepackage{booktabs} 
\usepackage{url}
\usepackage{doi}
\begin{document}

\title{Energy Transfer and Data Collection from Batteryless Sensors in Low-altitude Wireless Networks}

\author{Wen Zhang,
        Aimin Wang,
        Jiahui Li, \IEEEmembership{Member,~IEEE,}
        Geng~Sun, \IEEEmembership{Senior Member,~IEEE,}
        Jiacheng Wang,\\
        Weijie Yuan, \IEEEmembership{Senior Member,~IEEE,}
        and Dusit Niyato, \IEEEmembership{Fellow,~IEEE}

  \thanks{This study is supported in part by the National Natural Science Foundation of China (62272194, 62471200), and in part by the Science and Technology Development Plan Project of Jilin Province (20250101027JJ), and in part by the National Research Foundation, Singapore, and Infocomm Media Development Authority under its Future Communications Research \& Development Programme, Defence Science Organisation (DSO) National Laboratories under the AI Singapore Programme (FCP-NTU-RG-2022-010 and FCP-ASTAR-TG-2022-003), Singapore Ministry of Education (MOE) Tier 1 (RG87/22), and the NTU Centre for Computational Technologies in Finance (NTU-CCTF).  (\emph{Corresponding authors: Jiahui Li and Geng Sun.})\protect}

  \thanks{
  \par Wen Zhang is with the College of Computer Science and Technology, Jilin University, Changchun 130012, China (e-mail: wenzhang24@mails.jlu.edu.cn).

  \par Aimin Wang is with the College of Computer Science and Technology, Jilin University, Changchun 130012, China (e-mail: wangam@jlu.edu.cn).

  \par Jiahui Li is with the College of Computer Science and Technology, and also with the Key Laboratory of Symbolic Computation and Knowledge Engineering of Ministry of Education, Jilin University, Changchun 130012, China (e-mail: lijiahui@jlu.edu.cn).

  \par Geng Sun is with the College of Computer Science and Technology, Key Laboratory of Symbolic Computation and Knowledge Engineering of Ministry of Education, Jilin University, Changchun 130012, China, and also with the College of Computing and Data Science, Nanyang Technological University, Singapore 639798 (e-mail: sungeng@jlu.edu.cn).

  \par Jiacheng Wang is with the College of Computing and Data Science, Nanyang Technological University, Singapore 639798 (e-mail: jiacheng.wang@ntu.edu.sg).
  
  \par Weijie Yuan is with the School of Automation and Intelligent Manufacturing, Southern University of Science and Technology, Shenzhen 518055, China (e-mail: yuanwj@sustech.edu.cn).
  
  \par Dusit Niyato is with the College of Computing and Data Science, Nanyang Technological University, Singapore 639798 (e-mail: dniyato@ntu.edu.sg).

  }
}

\maketitle

\begin{abstract}
The integration of wireless power transfer (WPT) with Internet of Things (IoT) offers promising solutions for sensing applications, but faces significant challenges when deployed in hard-to-access areas such as high-temperature environments. In such extreme conditions, traditional fixed WPT infrastructure cannot be safely installed, and batteries rapidly degrade due to hardware failures. In this paper, we propose an uncrewed aerial vehicle (UAV)-assisted data collection and WPT framework for batteryless sensor (BLS) networks deployed in these challenging environments. Specifically, we consider a practical scenario where a UAV first transfers energy to BLS nodes via WPT, enabling these nodes to subsequently transmit their collected data to the UAV through orthogonal frequency-division multiple access (OFDMA). Then, we formulate a multi-objective optimization problem that aims to maximize the fair data collection volume while minimizing the UAV energy consumption through joint optimization of transmit power allocation and flight trajectory planning. Due to the non-convex nature and dynamic characteristics of this problem, conventional optimization methods prove inadequate. To address these challenges, we propose an enhanced soft actor-critic algorithm with parameter-free attention, prioritized experience replay, and value-based reward centering (SAC-PPV), thereby improving the exploration efficiency and learning stability of the algorithm in complex WPT scenarios. Simulation results demonstrate that the proposed approach consistently outperforms benchmark algorithms under various network configurations.

\end{abstract}


%
%
\section{Introduction}
\label{sec:introduction}

\par The rapid proliferation of the Internet of Things (IoT) has revolutionized data collection across diverse sectors, with billions of devices generating vast amounts of information to support critical decision-making processes. In general, IoT systems usually face persistent energy constraints, particularly in power-intensive sensing applications deployed in challenging environments. Wireless power transfer (WPT) technology has consequently emerged as a promising solution for dynamically replenishing energy to IoT devices without physical connections. However, conventional WPT-based IoT deployments encounter insurmountable challenges in extreme environments such as high-temperature zones, volcanic regions, or hazardous industrial areas, where fixed WPT infrastructure cannot be safely installed, conventional batteries rapidly degrade under harsh conditions, and ground-based energy transmitters suffer from severe signal attenuation and hardware failures \cite{Cai2023, Librino2025}.

\par These environmental limitations have driven the development of batteryless sensors (BLSs) that can harvest energy directly from radio frequency (RF) signals \cite{Zhu2022}. BLS systems offer compelling advantages including enhanced sustainability through elimination of battery replacement, substantially reduced maintenance requirements, and the ability to operate in environments where battery access is impractical or hazardous. These characteristics make BLSs particularly valuable for long-term deployments in remote or extreme locations, from structural health monitoring in disaster-prone areas to environmental sensing in radioactive sites \cite{Li2025}. Nevertheless, BLS deployments in harsh environments still face two fundamental challenges that must be addressed simultaneously, \textit{i.e.}, establishing reliable power delivery mechanisms to sustain continuous operations and enabling efficient data retrieval from sensors in remote and physically inaccessible locations.

\par In this case, low-altitude wireless networks (LAWNs) present a reasonable solution to these dual challenges by introducing mobility and adaptability into IoT infrastructures \cite{wei2025}. Within this paradigm, uncrewed aerial vehicles (UAVs) function as mobile platforms that dynamically interact with ground-based sensors, effectively overcoming the limitations of fixed infrastructure \cite{Xu2024}. By integrating WPT modules, UAVs serve as aerial energy transmitters that provide targeted power to BLSs through directional RF signal transmission while concurrently collecting the generated sensor data. This dual capability makes UAV-assisted BLS networks particularly valuable for applications across industrial monitoring, disaster response, environmental surveillance, and security systems~\cite{Yang2025, Yi2025}. The inherent mobility of UAVs enables strategic position optimization to maximize both energy transfer efficiency and data collection quality, creating a highly adaptive system capable of supporting continuous operations in otherwise inaccessible environments.

\par However, current research on UAV-assisted data collection (e.g., \cite{Li2023, Liu2022}) predominantly focuses on scenarios involving self-powered sensors, thereby overlooking the unique energy-dependent constraints of BLS networks where sensing and transmission capabilities depend entirely on harvested energy. Similarly, studies on WPT optimization typically consider static infrastructure or simplified mobility patterns that inadequately represent the complexity of aerial power delivery in dynamic environments. Furthermore, existing frameworks predominantly employ single-objective approaches that optimize either data collection efficiency or energy consumption in isolation \cite{Zheng2024, Lu2025}, whereas BLS networks in harsh environments necessitate a multi-objective framework that simultaneously balances power delivery efficiency, data collection quality, and UAV operational constraints. These critical limitations render existing solutions inadequate for effective BLS deployment in extreme environmental conditions.

\par Thus, we propose a novel framework that leverages UAVs as mobile platforms for simultaneous WPT and data collection in BLS networks. In the proposed framework, the UAV dynamically adjusts its position and transmission parameters to efficiently deliver power to BLS nodes, which subsequently utilize the harvested energy to transmit their collected data back to the UAV via uplink communications. However, realizing this framework requires addressing three interconnected technical challenges. \textit{First}, optimizing the transmit power allocation of UAV to enhance BLS uplink performance under time-varying channel conditions \cite{Perera2021}. \textit{Second}, planning efficient UAV trajectories with fairness considerations that effectively balance the conflicting objectives of maximizing fair data collection while minimizing UAV energy consumption. \textit{Finally}, managing inherent system uncertainties arising from variable channel conditions, diverse energy harvesting efficiencies, and dynamic sensor requirements. These challenges necessitate an innovative solution approach capable of real-time adaptation while simultaneously optimizing multiple competing objectives under complex operational constraints.

\par 
Therefore, this work introduces an innovative real-time optimization approach designed for UAV-assisted joint data collection and WPT system for BLS networks. The primary contributions of this study can be outlined as follows:

\begin{itemize}
    \item \textit{UAV-assisted Joint Data Collection and WPT System for BLS Networks}: We consider a UAV-assisted joint data collection and WPT system for BLS networks. Specifically, the UAV provides continuous energy support to BLSs through WPT, while a BLS utilizes the harvested energy to upload sensed data to the UAV. To the best of our knowledge, such a joint optimization of data collection and WPT system for BLS networks has not yet been investigated in the literature.
    
    \item \textit{Joint Optimization Problem Formulation for UAV Transmit Power and Trajectory}: For the proposed system, we formulate a joint optimization problem to cooperatively plan the transmit power and trajectory of the UAV. To ensure equitable service across all BLSs, we define the fair data collection volume as the product of Jain's fairness index and the total collected data, thereby maximizing this fair data collection metric while minimizing the UAV energy consumption. This problem is challenging to solve due to its high real-time and dynamic nature.
    
    \item \textit{Enhanced Deep Reinforcement Learning (DRL) Approach}: We propose a DRL-based approach, namely, soft actor-critic with parameter-free attention module, prioritized experience replay, and value-based reward centering (SAC-PPV), to solve the formulated optimization problem. Specifically, SAC-PPV introduces a parameter-free attention module (PFAM) to better extract complex state features, a prioritized experience replay (PER) module to improve sample utilization efficiency, and a mechanism of value-based reward centering (VRC) to maintain training stability and accelerate algorithm convergence.
    
    \item \textit{Simulation and Performance Analysis}: The simulation outcomes indicate that the proposed algorithm outperforms various baselines. Moreover, we find that the SAC-PPV algorithm improves fair data collection volume by 10.80\% with the same simulation scenario compared with the basic SAC algorithm. Furthermore, the results demonstrate that our developed method supports BLS networks of different scales and performs optimally among benchmark algorithms.
\end{itemize}

\par The subsequent sections of this manuscript are structured as follows. Section~\ref{sec:related_work} examines existing literature. Section~\ref{sec:System Model And Preliminaries} presents the system model. Section~\ref{sec:Problem Formulation And Analyses} formulates the optimization problem. Section~\ref{sec:Muti-objective DRL-based Method} introduces the enhanced DRL algorithm. Section~\ref{sec:Simulation And Analyses} discusses simulation results, and Section~\ref{sec:Conclusion} offers concluding remarks.

%
%
\section{Related Work}
\label{sec:related_work}

\par In this work, we aim to maximize the fair data collection volume of the system while minimizing the energy consumption of the UAV through joint optimization of transmit power allocation and flight trajectory planning. In the following, we present key related works to highlight the novelty of this work.

%
%
\subsection{UAV-assisted Data Collection and WPT} 
\par Due to the mobility and flexibility, UAVs have been widely applied in IoT to achieve efficient data collection. For instance, the authors in~\cite{Wan2024} studied a multi-UAV route planning system for time-dependent data collection and proposed a hybrid tabu search-variable neighborhood descent algorithm to maximize collected data volume. Additionally, the authors in~\cite{Xueqiang2024} investigated UAV-assisted data collection networks and developed AI-empowered intelligent search algorithms to minimize flight distances while ensuring full sensor coverage. Moreover, the authors in~\cite{Fu2024} examined UAV-enabled data collection with flexible collection points and formulated a multiobjective optimization approach to minimize both task completion time and energy consumption. Furthermore, the authors in~\cite{Fan2024} addressed UAV path planning for mobile IoT device data collection and proposed a proximal policy optimization-based algorithm to maximize collected data while ensuring energy efficiency and collision avoidance. However, these works consider that IoT devices have sufficient energy for communication, which is not realistic in many practical scenarios where devices operate under severe energy constraints.

\par In recent years, WPT in UAV-assisted IoT networks has attracted considerable attention due to its ability to extend the lifetime of energy-limited devices. For instance, the authors in~\cite{Qi2024} investigated a UAV-assisted mobile crowdsensing system empowered by WPT. The authors formulated an optimization problem to maximize the volume of collected data through joint optimization of compression rate, WPT time, uploading time, transmit power, and UAV trajectory. Moreover, the authors in~\cite{Wangxj2024} studied a wireless powered wearable network with UAV deployment and user scheduling. They considered the height of UAVs for WPT to maximize network throughput and minimize task completion time. Additionally, the authors in~\cite{Lin2024} examined a wireless powered communication network enabled by a UAV equipped with reconfigurable intelligent surface. They proposed a two-timescale active and passive beamforming framework to maximize ergodic throughput under outdated channel state information and non-linear energy harvesting conditions. Furthermore, the authors in~\cite{Kim2024} investigated a wireless powered communication network enabled by multiple UAVs. The research focused on joint optimization of scheduling, transmit power, and trajectory to maximize the minimum uplink throughput while meeting energy neutrality and UAV mobility constraints. 

\par However, most existing systems consider that IoT devices have stable power sources or rechargeable batteries, with insufficient attention paid to BLS scenarios. In BLS networks, devices rely entirely on harvested energy for operation, which introduces unique challenges in terms of intermittent power availability and requires specialized protocols for reliable data collection and transmission.

%
%
\subsection{Joint Optimization of UAV Transmit Power and Trajectory}
\par Currently, optimization problems in UAV-assisted data collection systems primarily involve two optimization variables that include data collection volume and UAV energy consumption. Optimization of the former can improve the quality of service and information accuracy for IoT applications, while simultaneous optimization of the latter can reduce energy consumption for energy-constrained UAVs. For instance, the authors in~\cite{Zhang2025} studied UAV-enabled data collection with dynamic constraints and optimized transmit power of ground nodes and UAV trajectory to maximize average transmission data rate while considering UAV dynamics through control-based methods. The authors in~\cite{WangRui2024} proposed a rechargeable UAV trajectory optimization scheme for persistent data collection and minimized completion time by balancing collection, flight, and recharging time using SCA and heuristic methods. Moreover, the authors in~\cite{Jia2024} investigated multi-UAV data collection and optimized flight trajectories, hovering positions, and speeds to minimize both maximum task completion time and maximum energy consumption using a multi-objective ant colony optimization algorithm.

\par However, these works did not consider fairness during UAV data collection and merely treated UAV energy consumption as a constraint rather than an optimization objective. The treatment of both fair data collection volume and UAV energy consumption as optimization objectives would be more beneficial because this approach not only ensures equitable service quality among distributed IoT devices but also guarantees optimal UAV energy efficiency and extends UAV operation time, thereby achieving more balanced performance.

%
%
\subsection{Optimization Methods for UAV-assisted Data Collection and WPT}
\par In IoT solutions, many conventional optimization algorithms have been used to solve optimization problems related to UAV-assisted data collection. For example, the authors in~\cite{Lu2025}  studied a UAV-assisted resource allocation scheme for industrial IoT and proposed a joint device access, bandwidth allocation, power control and speed optimization algorithm based on queue tail distribution to minimize energy consumption while ensuring reliable data collection. Additionally, the authors in~\cite{Tang2024}  investigated UAV-assisted data collection in wireless sensor networks with hybrid NOMA and proposed both an optimization-based algorithm using block coordinate descent and a low-complexity heuristic algorithm to maximize the minimum throughput. However, conventional optimization methods typically require accurate mathematical models and complete environment information to solve optimization problems, which are difficult to obtain in this dynamic UAV-assisted data collection environment.

\par In such situations, DRL algorithms are more suitable for solving these optimization problems. For instance, the authors in~\cite{Wan2024a} studied multi-UAV scheduling for disaster data collection with time-varying data value, and proposed an attention-based DRL method to optimize UAV routes and service time at each node. Additionally, the authors in~\cite{Huang2025} investigated energy-efficient multi-UAV collaborative reliable storage without edge assistance, and proposed a centralized training and decentralized execution DRL algorithm based on Actor-Critic to optimize replica placement while minimizing energy consumption. Furthermore, the authors in~\cite{WangBo2025} explored a multi-UAV-assisted joint mobile edge computing and data collection system, and developed a soft actor-critic based approach with two-phase matching-based association to minimize MEC latency and maximize collected data volume. However, these DRL-based works did not address the critical issues of feature attention allocation, experience importance, and reward signal stability, which limited their overall performance and system reliability.

\par In summary, different from the existing works, we consider utilizing UAV integrating WPT modules to provide BLS networks with directional RF signal transmission while concurrently collecting the generated sensor data. Based on this, we aim to design the transmit power allocation and flight trajectory planning of UAV to maximize the fair data collection volume while minimizing the energy consumption of UAV.

%
\begin{figure}[!t]
  \centering
  \includegraphics[width=3.5in]{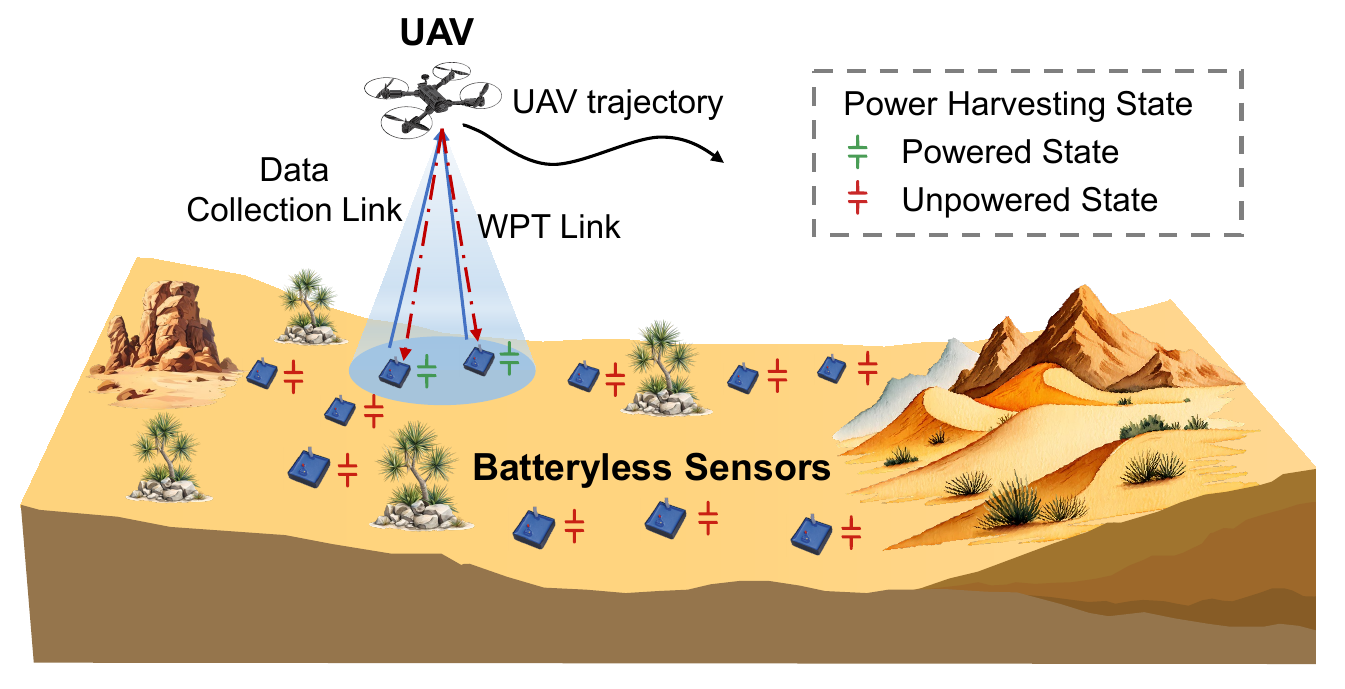}
  \caption{UAV-assisted joint data collection and WPT system for a BLS network. The UAV first charges BLSs within its distance threshold through WPT. Subsequently, the energized BLSs utilize the harvested energy to perform sensing and monitoring tasks, and then transmit the collected data back to the UAV via OFDMA.}
  \label{fig:network-model}
\end{figure}

%
%
\section{System Model And Preliminaries}
\label{sec:System Model And Preliminaries}

\par The following section outlines the system models and theoretical foundations utilized in our research. Specifically, we first present the network models, followed by the WPT model, the wireless data collection model, and the mobile and energy cost models of UAVs.

%
\begin{table}[t]
\caption{Main Notations}
\label{tab:notations}
\renewcommand{\arraystretch}{1.2}
\setlength{\tabcolsep}{3pt}
\small
\begin{tabular}{>{\raggedright\arraybackslash}p{0.25\columnwidth}>{\raggedright\arraybackslash}p{0.67\columnwidth}}
\Xhline{2\arrayrulewidth} 
\textbf{Notation} & \textbf{Definition} \\ 
\hline
$a_t^x$, $a_t^y$ & UAV flight distances along coordinate axes in time slot $t$ \\
$d_{i,t}^{US}$ & Distance between the UAV and the $i$-th sensor node in time slot $t$ \\
$d_0$ & Drag ratio \\
$D^{max}$ & Maximum coverage distance threshold for data collection and WPT by the UAV \\
$D_{i,t}$ & Data volume gathered by UAV from the $i$-th BLS in time slot $t$ \\
$E_t$ & Energy consumption of UAV in time slot $t$ \\
$G$ & Rotor disc area \\
$h_{i,t}$ & Communication channel gain between UAV and the $i$-th BLS in time period $t$ \\
$N_{SN}$ & Number of BLSs \\
$P_t^U$ & Transmit power of UAV in time slot $t$ \\
$P_{i,t}^R$ & RF power received by the $i$-th BLS in time slot $t$ \\
$P_{i,t}^H$ & Harvested DC power converted from the received RF power at the $i$-th BLS in time slot $t$ \\
$R_{i,t}$ & Achievable rate between BLS $i$ and UAV in time slot $t$ \\
$s_0$ & The rotor solidity \\
$t$, $T$, $\mathcal{T}$ & Time slot index, total count, and collection respectively \\
$t_d$ & Length of each time slot \\
$U_{tip}$ & Blade tip speed of the rotor \\
$V$ & Average induced airflow velocity during hovering \\
$x^{SN}_i$, $y^{SN}_i$, $z^{SN}_i$ & Position coordinates of the $i$-th BLS \\
$x_t^{U}$, $y_t^{U}$, $z_t^{U}$ & UAV position coordinates during time interval $t$ \\
$X^{min}$, $X^{max}$ & Minimum and maximum boundaries of operational region along x-axis \\
$Y^{min}$, $Y^{max}$ & Lower and upper limits of target zone along y-axis \\

$\beta_{i,t}$ & Large-scale fading effects in time slot $t$ \\
$\rho_0$ & Air density \\
$\sigma^2$ & Noise power \\

\Xhline{2\arrayrulewidth} 
\end{tabular}
\end{table}

%
\subsection{Network Segments}
\par As shown in Fig.~\ref{fig:network-model}, a UAV-assisted joint data collection and WPT system is deployed to operate in remote areas, such as a desert environment. Specifically, the monitoring area $A_{m}$ contains a large number of sensors. Due to the extreme temperature fluctuations and frequent sandstorms in the desert, conventional battery-powered sensors are easily to be damaged, and they are challenging to maintain~\cite{Cai2023}. To address these challenges, a group of BLSs equipped with high-efficiency capacitors is adopted for data acquisition and environmental monitoring because these capacitors offer superior environmental tolerance and rapid charge-discharge capabilities. These sensors are represented by the set $\mathcal{W} = \{ i \mid i = 1, 2, \ldots, N_{SN} \}$. Given the challenges of installing fixed power infrastructure and base stations (BSs) in desert areas, a quadrotor UAV is deployed to first deliver power to the BLSs via WPT and subsequently collect the transmitted data. Specifically, each sensor and UAV operate with an omnidirectional antenna, while the positions of the BLSs are predetermined and static. Moreover, the range of the UAV for data collection and WPT is limited by the distance threshold $D_{max}$.

\par In each mission cycle, the BLSs need to be powered and upload their sensed data for data backup and security. For clarity, time is partitioned into $T$ slots, each with a duration of $t_d$, and denoted by $\mathcal{T} = \{ t \mid t = 1, 2, \ldots, T \}$ \cite{Bakhrani2025}. Each BLS has a certain probability of generating new data collection requests in each time slot. The heterogeneous spatial distribution of BLSs, coupled with the dynamic nature of data requests and the variable energy consumption patterns of the UAV, poses significant challenges to real-time trajectory optimization. To address these challenges, the UAV dynamically adjusts its flight path to simultaneously engage with multiple BLSs within a single time slot, thereby coordinating WPT and data transmission. Specifically, the UAV performs downlink WPT to replenish BLS energy storages, while employing an orthogonal frequency-division multiple access (OFDMA) scheme for bandwidth allocation during uplink data transmission \cite{Xuhui2025}.

\par Considering a three-dimensional (3D) Cartesian coordinate system, the position of the $i$th sensor in the monitoring area $A_{m}$ is represented as $\mathbf{c}_i^{SN} = (x_i^{SN}, y_i^{SN}, 0)$. Since the duration of each time slot $t_d$ is sufficiently small, the position of UAV is regarded as fixed within any time slot $t$ and is expressed as $\mathbf{c}^{U}_t \triangleq (x^U_t, y^U_t, z^U), \forall t \in \mathcal{T}$. To maintain balanced data acquisition in the BLS network and optimize UAV energy consumption, we subsequently formulate the key operational models, including the WPT model, wireless data collection model, and UAV mobile and energy cost models.

%
%
\subsection{WPT Model}
\par During time slot $t$, we consider the UAV to be capable of wirelessly powering the BLS nodes within its coverage area using a transmit power of $P^U_t$. Let $d_{i,t}^{US}$ represent the distance between the UAV and the $i$-th sensor node during time slot $t$. When the distance between sensor $i$ and the UAV is within the threshold $D_{\text{max}}$, sensor $i$ can function properly. Then, the power received by sensor $i$ can be expressed using the Friis transmission equation as follows~\cite{Panahi2024}:
\begin{equation}
P^R_{i,t} = P^{U}_t \frac{G_T G_R \lambda^2}{(4 \pi)^2 (d_{i,t}^{US})^\alpha},
\end{equation} 
where $\lambda$ is the wavelength of the RF signal, $\alpha$ is the path loss exponent of the air-to-ground channel, and $G_T$ and $G_R$ are the antenna gains of the UAV and the BLS, respectively. Consider that the minimum power required to activate the rectifying antenna is $P_{\mathrm{min}}$, and $P_{\mathrm{max}}$ represents the saturation limit of the received power of the sensor. Then, at a certain time slot $t$, the most precise energy harvesting model for the received power of the BLS $i$ can be represented as follows \cite{Cetinkaya2023}:
\begin{equation}
P^H_{i,t}(P^R_{i,t}) =
\begin{cases}
    0, & P_{i,t}^R \in [0, P_{\mathrm{min}}), \\
    \eta(P_{i,t}^R) \cdot P_{i,t}^R, & P_{i,t}^R \in [P_{\mathrm{min}}, P_{\mathrm{max}}), \\
    P_{i,t}^H (P_{\mathrm{max}}), & P_{i,t}^R \geq P_{\mathrm{max}},
\end{cases}
\end{equation}  
where $P^H_{i,t}$ denotes the transformation of the received power $P_{i,t}^R$ into available direct current power. Through the application of curve-fitting techniques to experimental data, the function $\eta(\cdot)$ can be approximated as a polynomial form, which yields a computationally manageable expression while maintaining adequate precision \cite{Cetinkaya2023}. In this work, it is considered that the rectifying antenna operates within an ideal range (i.e., $P_{i,t}^R \in [P_{\mathrm{min}}, P_{\mathrm{max}}]$).

\par After completing the WPT, we proceed to discuss the communication model for data collection from the BLS by the UAV.

%
%
\subsection{Wireless Data Collection Model}
\par The scenario where the UAV collects data from BLSs is considered as a typical ground-to-air communication model. The channel coefficient $h_{i,t}$ between UAV and BLS $i$ at time slot $t$ is given by \cite{Wan2024}
\begin{equation}
h_{i,t}=\sqrt{\beta_{i,t}} \tilde{h}_{i,t},
\end{equation}
\noindent where $\beta_{i,t}$ characterizes the large-scale fading phenomena, while $\tilde{h}_{i,t}$ denotes the small-scale fading component as a complex random variable satisfying $\mathbb{E}[|\tilde{h}_{i,t}|^2] = 1$. In general, an elevation angle-dependent probabilistic probabilistic line-of-sight (LoS) model is adopted. This is due to the fact that the UAV can try to achieve a LoS link even in the presence of obstacles by moving. Under the proposed model, large-scale attenuation is conventionally treated as a random process contingent upon LoS and non-LoS (NLoS) occurrence probabilities. Accordingly, the large-scale channel coefficient $\beta_{i,t}$ is defined as follows:\cite{Wan2024}:
\begin{equation}
{\beta }_{i,t}= \begin{cases}
    \beta _{0} ({d_{i,t}^{US}}) ^{-\alpha}, & \text{LoS environment}, \\
    \kappa \beta _{0} ({d_{i,t}^{US}}) ^{-\alpha }, & \text{NLoS environment},
\end{cases}
\end{equation}
\noindent where $\beta_0$ corresponds to the path loss at reference distance, $\alpha$ denotes the path loss exponent, and $\kappa$ signifies the supplementary fading component attributed to NLoS environments. Additionally, the LoS probability is characterized by a logistic function dependent on elevation angle $\theta_{i,t}$, formulated as follows \cite{Wan2024}:
\begin{equation}
P_{i,t}^{ \text{LoS}}=\frac{1}{1+C\exp(-D[\theta_{i,t}-C])},
\end{equation}
where the parameters $C$ and $D$ represent environment-specific modeling coefficients, and the elevation angle in degrees is calculated as $\theta_{i,t} = 180/\pi \sin^{-1} ( z^U/d_{i,t}^{US} )$. Moreover, we define the blocked propagation probability as $P_{i,t}^{\text{NLoS}} = 1 - P_{i,t}^{\text{LoS}}$. Therefore, the expected channel gain obtained by averaging over both random effects is expressed as follows:
\begin{equation}
\mathbb{E}[{|{h}_{i,t}|}^{2}] = P_{i,t}^{\text{LoS}} \beta _{0}({d_{i,t}^{US}})^{-\alpha} + P_{i, t}^{\text{NLoS}} \kappa \beta _{0}({d_{i,t}^{US}})^{-\alpha}.
\end{equation}

\par The achievable rate between BLS $i$ and the UAV during time slot $t$, expressed in bits per second (bps), can be expressed as follows \cite{Pan2023}:
\begin{equation}
R_{i,t}=B_{i,t}\log_{2}\left(1+\frac{P_{i,t}^{H}{|{h}_{i,t}|}^{2}}{\sigma^{2}}\right),
\end{equation}
where $\sigma^2$ denotes the noise power at the receiver, and $B_{i,t}$ represents the communication bandwidth allocated to the BLS $i$ during the time slot $t$. Consequently, the volume of data collected by the UAV from the $i$-th BLS during time slot $t$ can be calculated as follows:
\begin{equation}
D_{i,t} = R_{i,t} \cdot t_d,
\end{equation}
where $t_d$ represents the duration of the data transmission period within each time slot.

\par In this work, the UAV possesses the capability to adjust its position for achieving better wireless data collection efficiency. Due to this mobility feature, we need to carefully consider both the movement capabilities and the corresponding energy consumption patterns of the UAV. Therefore, in the following sections, we provide detailed descriptions of the mobility model and energy consumption model of the UAV to complete our system formulation.

%
%
\subsection{UAV Mobility and Energy Consumption Models}
\par In time slot $t$, the movement of the UAV is determined by two horizontal displacements, executed through the action $\mathbf{a}_t = [a^x_t, a^y_t, 0]$. Therefore, the position of the UAV during time slot $t$ is updated as follows:
\begin{equation}
\mathbf{c}^{U}_t = \mathbf{c}^{U}_{t-1} + \mathbf{a}_t.
\end{equation}

\par Then, for the propulsion energy consumption of the UAV, the commonly used model from reference~\cite{Jun2025} is adopted, which is given by
\begin{equation}
\begin{aligned}
E^P_t = &\left( P_0\left(1 + 3\left(\frac{v_t}{U_{tip}}\right)^2\right) + P_m\left(\sqrt{1 + \frac{1}{4}\left(\frac{v_t}{V}\right)^4} \right. \right. \\
&\left. \left. - \frac{1}{2}\left(\frac{v_t}{V}\right)^2\right)^{\frac{1}{2}} + \frac{1}{2} d_0 \rho_0 s_0 G \left(v_t\right)^3\right) t_d,
\end{aligned}
\end{equation}
where $U_{tip}$ and $V$ denote the tip speed of the rotor blade and the mean rotor-induced velocity during hovering, respectively. Additionally, $d_0$, $\rho_0$, $s_0$, and $G$ represent the drag coefficient of the UAV body, air density, rotor solidity, and rotor disc area, respectively. Moreover, $P_0$ and $P_m$ are two constants representing the rotor power and the induced power during hovering, corresponding to the power consumption of the rotor. The variable $v_t=\sqrt{(a^x_t)^2+(a^y_t)^2} / t_d$ denotes the speed of the UAV in time slot $t$.

\par In summary, the energy consumption of UAV consists of charging and propulsion energy consumption in time slot $t$ can be given by
\begin{equation}
E_t = \left(E^C_t + E^P_t\right).
\end{equation}

%
%
\section{Problem Formulation and Analyses}
\label{sec:Problem Formulation And Analyses}

\par This section focuses on formulating an optimization problem to enhance data collection process from ground sensor networks. We begin by identifying key system considerations, followed by defining decision variables and optimization objectives. Finally, we establish a multi-objective optimization problem with corresponding analytical treatment.

\par At any time slot $t \in \mathcal{T}$, the transmit power and spatial position of UAV jointly determine both the data collection performance and energy consumption. As such, the trajectory planning and power allocation demonstrate interrelated characteristics with inherent coupling. Simultaneously, the positioning decisions of UAV affect the fairness of data collection among sensors, while the power allocation impacts both the charging efficiency and flight energy consumption. Consequently, these optimization objectives possess mutually conflicting associations. To address the fairness concern, we utilize Jain's fairness index that considers long-term fairness rather than instantaneous quality of service metrics \cite{Tarekegn2025}, as UAV-based services typically operate over extended periods where optimal solutions may include temporary partial service phases. Therefore, the interconnected nature of parameters and reciprocal impact among goals necessitate a multi-criteria optimization approach. The decision variables are defined below.

\par The decision variables are established and jointly optimized as follows: (i) $\mathbf{P}=\left\{P_{t}^U \mid t \in \mathcal{T}\right\}$, which is a continuous variable matrix that specifies UAV transmit power across time slots for BLS energy transfer. (ii) $\mathbf{A}=\left\{\left[a^x_t, a^y_t, 0\right] \mid t \in \mathcal{T}\right\}$, a matrix consisting of continuous variables that defines the spatial displacement of the UAV over time for trajectory optimization. Next, we present the formulation of the optimization objectives under consideration.

\par The primary objective is to maximize the fair data collection volume from BLSs over the total timeline. To ensure fairness, we incorporate Jain's fairness index given by
\begin{equation}
J\left(D_{i,t}\right)=\frac{\left(\sum_{i \in \mathcal{W}}D_{i,t}\right)^2}
{N_{S N} \left(\sum_{i \in \mathcal{W}}D_{i,t}^2\right)},
\end{equation}
where $J\left(D_{i,t}\right)\in [0, 1]$ and a higher fairness index indicates more equal data collection distribution across all sensors. As such, our first optimization objective is given by
\begin{equation}
f_1 = \sum_{t \in \mathcal{T}}\sum_{i \in \mathcal{W}}D_{i,t}J\left(D_{i,t}\right).
\end{equation}

\par Additionally, when the UAV performs data collection and BLS charging operations, both flight maneuvers and WPT contribute to energy consumption. Due to the constrained onboard energy resources of UAVs, the second objective function aims to minimize the aggregate energy consumption of the UAV, which is given by
\begin{equation}
f_2 = \sum_{t \in \mathcal{T}}E_t.
\end{equation}

\par According to the two optimization objectives above, our optimization problem can be formulated as follows:
\begin{subequations} \label{eq:joint_opt}
\begin{align}
\min _{\mathbf{P},\mathbf{A}} 
& \quad F = \{f_1,-f_2\}, \\
\text{s.t.} 
& \quad C1: X_{\min} \leq x^U_t \leq X_{\max}, \quad \forall t, \label{eq:C1} \\
& \quad C2: Y_{\min} \leq y^U_t \leq Y_{\max}, \quad \forall t, \label{eq:C2} \\
& \quad C3: \left\|a^x_t\right\| \leq x_{\max}, \quad \forall t, \label{eq:C3} \\
& \quad C4: \left\|a^y_t\right\| \leq y_{\max}, \quad \forall t, \label{eq:C4} \\
& \quad C5: P_\mathrm{min}^{U} \leq P_t^{U} \leq P_\mathrm{max}^{U}, \quad \forall t, \label{eq:C5} \\
& \quad C6: R_{i,t} \geq R_{th}, \quad \forall t, \label{eq:C6}
\end{align}
\end{subequations}
where (\ref{eq:C1}) and (\ref{eq:C2}) indicate that the UAV can only fly within the designated target area. Moreover, the movement distance of the UAV in each time slot is constrained by (\ref{eq:C3}) and (\ref{eq:C4}). In addition, (\ref{eq:C5}) limits the transmit power for charging the UAV. The parameter $R_{th}$ represents the minimum achievable rate. Furthermore, (\ref{eq:C6}) specifies that the communication rate must satisfy the minimum rate constraint $R_{th}$. As noted in \cite{He2023, Zhendong2025}, this problem is non-convex and generally quite difficult to solve in real-time.

%
%
\section{Proposed SAC-PPV Algorithm} 
\label{sec:Muti-objective DRL-based Method}
\par 
This section presents a DRL-based approach for addressing the joint optimization challenge. Initially, we demonstrate the motivations for using DRL, and transform the optimization challenge into a Markov decision process (MDP) formulation. Subsequently, we introduce the soft actor-critic (SAC) algorithm and its fundamental principles. Finally, we detail the SAC-PPV algorithm and its implementation framework.

%
%
\subsection{Motivations for Using DRL and MDP Framework}
\label{ssec:Motivations for Using DRL and MDP Formulation}

\par 
The problem exhibits dynamism and uncertainty characteristics due to the unpredictability of factors such as sensor data transmission demands and UAV mobility, which can change rapidly. As a result, conventional static optimization methods, including convex and non-convex optimization~\cite{Li2024}, are not effective for handling this complexity. In contrast, DRL offers the ability to quickly adapt to changing conditions and provides reliable solutions. Therefore, we employ the DRL approach to tackle our optimization challenge.

\par We transform the joint optimization challenge presented in Eq. \eqref{eq:joint_opt} into an MDP decision process framework. From a mathematical perspective, the MDP framework consists of a five-tuple representation $(\mathcal{S}, \mathcal{A}, \mathcal{P}, R, \gamma)$. Here, $\mathcal{S}$ represents the state space, while $\mathcal{A}$ denotes the action space. Additionally, $\mathcal{P}$ corresponds to state transition probabilities, $R$ defines the reward function, and $\gamma$ indicates the discount factor. Furthermore, state, action, and reward represent the three fundamental elements that constitute the core components of this formulation. Subsequently, we provide detailed descriptions of these essential components.

\begin{enumerate}[label=\textbf{1)}, left=0pt]
    \item \textbf{State Space:} We consider that the UAV maintains precise positioning capabilities and can acquire real-time information about its operational environment. Concurrently, the system possesses complete knowledge of BLS locations within the target area. Accordingly, the state space of the MDP incorporates these essential and observable conditions under which the UAV-assisted BLS network operates. Consequently, the state at time slot $t$ is defined as follows:
    \begin{align}
    s_t=\left\{\mathbf{c}^U_t,\mathbf{c}^{SN} \right\},
    \end{align}
    where $\mathbf{c}^U_t$ denotes the current position of the UAV, and $\mathbf{c}^{SN} \triangleq \{\mathbf{c}_i^{SN}\},\forall i \in \mathcal{W}$ represents the coordinates of all BLSs.
\end{enumerate}
\begin{enumerate}[label=\textbf{2)}, left=0pt]
    \item \textbf{Action Space:} As previously established, the optimization requires joint determination of UAV transmit power and trajectory to maximize fair data collection while minimizing energy consumption of the UAV. Moreover, the UAV is required to make real-time decisions on both its spatial displacement and transmit power at each time slot to adapt to the dynamic sensor network environment. Therefore, the possible actions at time slot $t$ for the UAV are defined as follows:
    \begin{align}
    a_t=\left(a^x_t, a^y_t, P_{t}^U\right),
    \end{align}
    where $a^x_t, a^y_t$ are the flight distance of the UAV along the $x$ and $y$ axes. Moreover, $P_{t}^U$ is the transmit power of the UAV in time slot $t$.
\end{enumerate}
\begin{enumerate}[label=\textbf{3)}, left=0pt]
    \item \textbf{Reward Function:} In a DRL-based framework, the agent depends on the reward to modify its actions and develop optimal policies for maximizing the reward. Therefore, the design of the reward system is crucial in enhancing the performance of the system. In this study, we aim to maximize fair data collection volume while minimizing the energy consumption of the UAV. To achieve long-term optimization of both objectives, the reward $r_t \in \mathcal{R}$ is defined as follows:
    \begin{align}
    r_t = \frac {\xi\sum_{t'=1}^t\sum_{i \in \mathcal{W}}D_{i,t'}J\left(D_{i,t'}\right)}  {\sum_{t'=1}^t E_{t'}} - r^P_t, \label{eq:reward}
\end{align}
    where $r^P_t$ is the penalty when the UAV flies out of the target area. The parameter $\xi$ is weighting factor that control the trade-off between fair data collection performance and UAV energy consumption.
\end{enumerate}

%
%
\subsection{SAC Algorithm}
\label{ssec:SAC}
\par
We introduce the SAC algorithm~\cite{Wang2024} as the framework to handle the MDP. Specifically, SAC aims to maximize the expected reward and also maximize the entropy of a policy to encourage exploration and guarantee robust performance under uncertain environments. The primary objective of SAC involves finding the optimal policy that maximizes the expected cumulative reward along with an entropy term. This objective is given by
\begin{align}
\pi^*=\underset{\pi}{\operatorname{argmax}} E_{\left(s_t, a_t\right) \sim \rho_\pi}\left[\sum_t R\left(s_t, a_t\right)+\alpha H\left(\pi\left(\cdot \mid s_t\right)\right)\right],
\label{eq:sac_objective}
\end{align}
where $\pi^*$ represents the optimal policy, $\rho_\pi$ denotes the state-action marginal distribution induced by policy $\pi$, $R(s_t, a_t)$ is the reward function, $\alpha$ is a temperature parameter that balances reward maximization against entropy maximization, and $H(\pi(\cdot|s_t))$ is the entropy of the policy at state $s_t$.
The entropy term $H(P)$ is given by
\begin{align}
H(P)=\underset{x \sim P}{\mathbb{E}}[-\log P(x)],
\label{eq:entropy_objective}
\end{align}
where $P$ represents a probability distribution, and $x$ is a random variable sampled from $P$.
To optimize this objective, SAC utilizes a soft Q-function that incorporates the entropy term. Specifically, the soft Q-function is given by
\begin{align}
Q\left(s_t, a_t\right) = r\left(s_t, a_t\right) + \gamma \cdot V_{t+1},
\end{align}
where $r(s_t, a_t)$ is the immediate reward, $\gamma$ is the discount factor, and $V_{t+1} = E_{s_{t+1}, a_{t+1}}[Q(s_{t+1}, a_{t+1})-\alpha \log (\pi(a_{t+1} \mid s_{t+1}))]$ represents the expected future value that incorporates both the Q-value and policy entropy at the next state $s_{t+1}$. This formulation clearly separates the immediate reward from the discounted future return with entropy regularization.
\par
In practice, SAC alternates between updating the Q-function parameters and the policy parameters. The Q-function parameter update minimizes the mean squared Bellman error. The formula for this update is given by
\begin{align}
J_Q(\theta)=E_{\left(s_t, a_t, s_{t+1}\right) \sim D, a_{t+1} \sim \pi_{\varphi}}\left[\frac{1}{2}\left(Q_\theta\left(s_t, a_t\right)- y_t \right)^2\right],
\end{align}
where $D$ represents the replay buffer, $\theta$ are the Q-function parameters, and $y_t$ is the target value defined as $y_t = r_t+\gamma\left(Q_\theta^{\prime}\left(s_{t+1}, a_{t+1}\right)-\alpha \log \left(\pi_{\varphi}\left(a_{t+1} \mid s_{t+1}\right)\right)\right)$, where $\theta'$ are the target Q-function parameters and $\varphi$ are the policy parameters.
\par
Additionally, the policy parameter update minimizes the expected Kullback-Leibler divergence. The objective is given by
\begin{equation}
\begin{aligned}
J_\pi(\varphi) &= \mathbb{E}_{s_t \sim D, a_t \sim \pi_{\varphi}}\Big[ 
\log \pi_{\varphi}(a_t \mid s_t) - \frac{1}{\alpha} Q_\theta(s_t, a_t)\\
 &\quad+ \log Z(s_t) \Big].
\end{aligned}
\end{equation}
Furthermore, the temperature parameter $\alpha$ can also be optimized to achieve a desired level of entropy. The objective for $\alpha$ is given by
\begin{align}
J(\alpha)=\mathbb{E}_{a_t \sim \pi_t}\left[-\alpha \log \pi_t\left(a_t \mid s_t\right)-\alpha H_0\right],
\end{align}
where $H_0$ is a target entropy value.
To enable efficient computation of policy gradients through backpropagation, SAC employs the reparameterization trick. The formula for action sampling is given by
\begin{align}
a_t=f_{\varphi}\left(\varepsilon_t ; s_t\right)=f_{\varphi}^\mu\left(s_t\right)+\varepsilon_t \odot f_{\varphi}^\sigma\left(s_t\right),
\end{align}
where $\varepsilon_t$ is a noise vector sampled from a standard normal distribution, $f_{\varphi}^\mu(s_t)$ represents the mean of the policy distribution, $f_{\varphi}^\sigma(s_t)$ represents the standard deviation, and $\odot$ denotes element-wise multiplication.
\par
Through iterative updates of the Q-function, policy, and temperature parameters, SAC converges to a policy that effectively balances exploitation and exploration. This balance results in robust performance across a variety of reinforcement learning tasks. However, SAC faces challenges in high-dimensional state and action spaces due to the difficulty of modeling complex relationships and the inefficiency in sample utilization. Therefore, these limitations motivate us to enhance SAC for improved performance in complex environments.

\begin{figure*}
  \centering
  \includegraphics[width=0.9\textwidth]{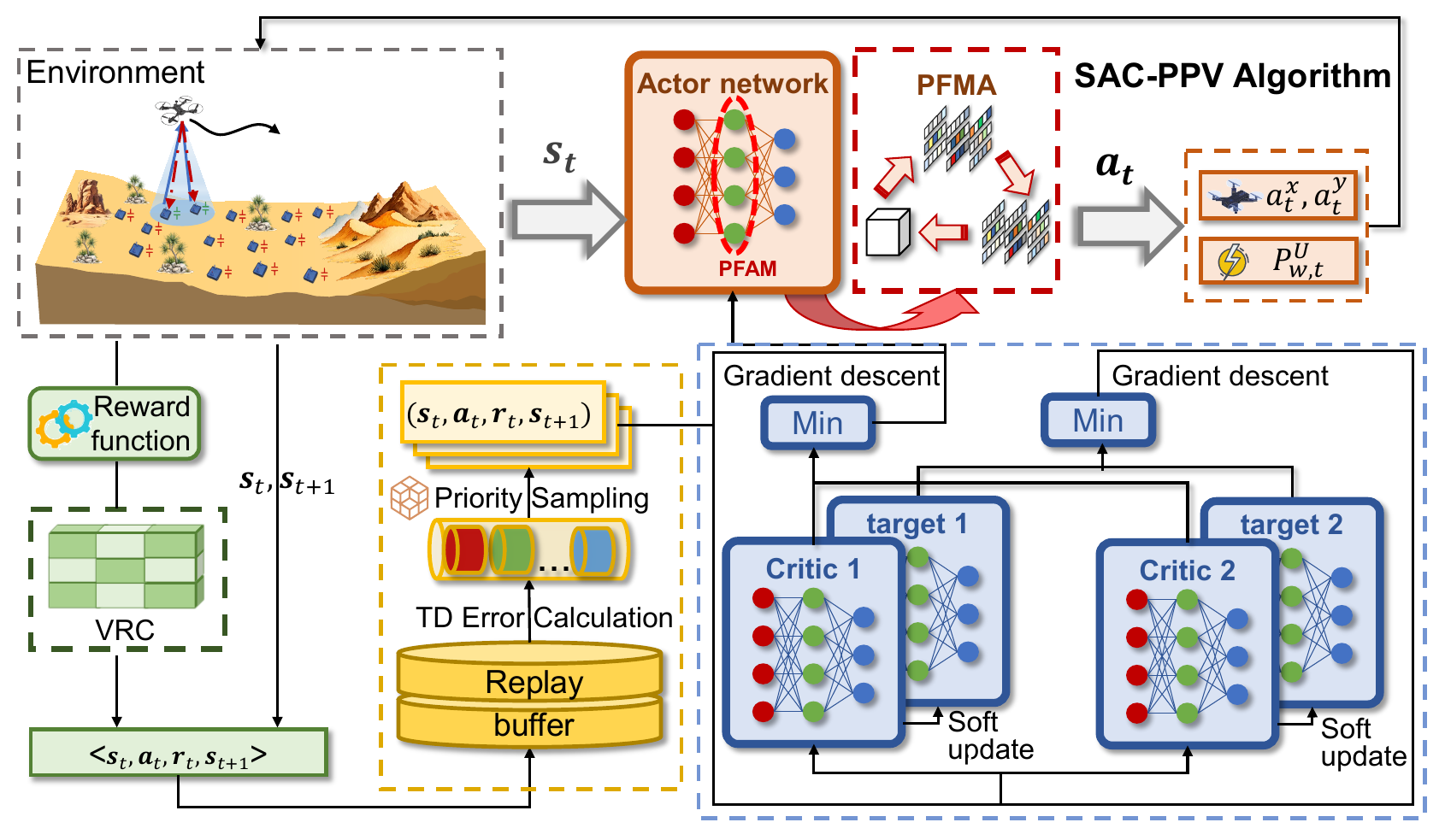}
  \caption{The structure of the proposed SAC-PPV algorithm integrates three key improvements to the standard SAC framework. PFAM is integrated into the actor network to enhance the perception capability of the policy for key environmental features. Moreover, the replay buffer employs PER to improve sample efficiency and learning speed. In addition, VRC is applied after the reward function to enhance training stability.}
  \label{fig:SAC-PPV}
\end{figure*}

\subsection{SAC-PPV}
\label{ssec:SAC-PPV}

\par In this section, we introduce an enhanced version of the SAC algorithm, namely SAC-PPV. The dynamic nature of the state space creates convergence difficulties for conventional SAC algorithms because the agent cannot effectively identify and focus on critical environmental features during the learning process. Moreover, this difficulty is further exacerbated by the uniform sampling mechanism employed in the replay buffer of SAC. This mechanism treats all experiences with equal importance regardless of their actual learning potential, thereby compromising the efficiency of UAV data collection and power transmission optimization. Additionally, the high variance inherent in reward signals leads to training instability, making it considerably more challenging for the algorithm to learn a reliable and consistent policy.

\par To address these fundamental limitations, SAC-PPV integrates three key improvements that work synergistically to improve algorithm performance by enhancing feature perception, improving sample efficiency, and stabilizing the training process. The detailed implementation and theoretical foundations of each improvement are presented in the following subsections.

\subsubsection{Parameter-Free Attention Module (PFAM)} The standard SAC algorithm often has limitations in feature extraction capabilities when processing complex state spaces. This deficiency leads to suboptimal policy learning and inefficient exploration in environments with high-dimensional or noisy state representations. Therefore, to enhance the feature recognition ability of algorithm when processing complex state spaces, we first introduce PFAM into the SAC algorithm \cite{Yang2021}. PFAM differs from previous approaches because it calculates neuron importance directly rather than handling channel and spatial attention separately. Specifically, the algorithm directly calculates the importance of each neuron based on its distinctiveness from surrounding neurons. Moreover, the neuron importance estimation process evaluates how separable each neuron is from its context using our energy function that requires no learnable parameters. As such, the energy function is given by
\begin{align}
e_t^* = \frac{4(\hat{\sigma}^2 + \omega)}{(t - \hat{\mu})^2 + 2\hat{\sigma}^2 + 2\omega},
\end{align}
where $\hat{\mu}$ and $\hat{\sigma}^2$ represent the mean and variance of the features, respectively, and $\omega$ is the regularization parameter. Lower energy values indicate that the neuron shows more distinctive patterns compared to surrounding neurons. Furthermore, the final step refines the original features through element-wise multiplication with the sigmoid-normalized importance weights. This refinement process is given by
\begin{align}
\tilde{\mathbf{X}} = \mathrm{sigmoid}(\frac{1}{\mathbf{E}}) \odot \mathbf{X},
\end{align}
where $\mathbf{E}$ represents the energy values for all neurons, and $\tilde{\mathbf{X}}$ is the refined feature representation. This simple yet effective approach allows PFAM to achieve superior performance with minimal computational overhead. Based on these advantages, we integrate PFAM between hidden layers of the actor network in our proposed SAC-PPV algorithm. Through adaptive weighting of state representations, the policy network can capture critical state changes in the environment more precisely, thus making more optimal decisions.

\subsubsection{Prioritized Experience Replay (PER)} In the standard SAC algorithm, samples are selected randomly from the experience pool, which may overlook the importance of certain state transitions. Moreover, due to the limited capacity of the experience pool, valuable experiences may be replaced before being fully utilized, further destabilizing the learning process. To address this issue, we introduce PER \cite{Li2024a}. Specifically, PER assigns higher sampling probabilities to experiences with large temporal difference (TD) errors, enabling the algorithm to focus on critical experiences and enhance learning efficiency. As such, the TD error is computed as follows:
\begin{align}
\delta_i = \left| r_t + \gamma Q(s_{t+1}, a') - Q(s_t, a_t) \right|,
\end{align}
where $\delta_i$ reflects the value of updating the current policy. Following this, the probability of selecting a sample is calculated as follows:  
\begin{align}
P(i) = \frac{p_i^\alpha}{\sum_j p_j^\alpha}.
\end{align}
To prevent bias and ensure stability, importance sampling weights are applied during gradient updates, \textit{i.e.,}  
\begin{align}
w_i = \left( \frac{1}{N} \cdot \frac{1}{P(i)} \right)^\beta.
\end{align}
\par By applying PER, SAC-PPV retains more valuable data for policy learning, which helps reduce bias during training. This improvement leads to faster convergence and more efficient data utilization.

\subsubsection{Value-Based Reward Centering (VRC)} To address the inefficiency and convergence difficulties caused by unstable reward signals, we introduce VRC \cite{Naik2024}. Unlike reward centralization methods that use simple running averages, VRC integrates TD errors with average reward estimation to provide more accurate reward centralization. The formula for conventional reward centralization to calculate the running average is given by
\begin{align}
\bar{r}_{t+1} = \bar{r}_t + \beta_t(r_{t+1} - \bar{r}_t),
\end{align}
where $\bar{r}_t$ represents the estimated average reward of target policy $\pi$ at time step $t$, and $\beta_t$ represents the learning rate for average reward estimation at time step $t$. However, this method leads to convergence to the average reward of the behavior policy rather than the target policy. This results in biased estimates that negatively affect algorithm convergence and learning efficiency.
Therefore, we adopt the value-based reward centralization approach. This method updates the average reward estimate using TD errors. The update formula is given by
\begin{align}
\bar{r}_{t+1} = \bar{r}_t + \eta \alpha_t \rho_t \delta_t,
\end{align}
where $\delta_t$ represents the TD error, $\rho_t = \pi(a_t|s_t) / b(a_t|s_t)$ represents the importance sampling ratio, $\alpha_t$ represents the learning rate, and $\eta$ represents the average reward update rate. This approach connects average reward estimation and value estimation in an interdependent manner, which provides more accurate estimates \cite{Naik2024}.
\par
In the SAC algorithm, we focus on relative changes in rewards rather than absolute values by subtracting the estimated average reward from immediate rewards. We combine value-based reward centralization with the original target Q-value calculation and modify the target Q-value formula as follows:
\begin{align}
Q_{target} =\;& (r_t - \bar{r}_t) + \gamma \cdot (1 - D_t) \cdot 
\big(Q_{min}(s_{t+1}, a_{t+1}) 
\nonumber\\
&- \alpha \cdot \log\pi(a_{t+1}|s_{t+1})\big),
\end{align}
where $\gamma$ represents the discount factor, $D_t$ indicates terminal state, $Q_{min}$ represents the minimum value of the two Q networks, $\alpha$ represents the temperature parameter, and $\log\pi$ represents the log probability of the policy. By subtracting the estimated average reward from immediate rewards, the algorithm can better focus on relative changes in rewards rather than absolute values. As a result, this approach significantly improves learning efficiency.

\par \textit{3) Main Steps of SAC-PPV Algorithm:} Figure~\ref{fig:SAC-PPV} illustrates the framework of SAC-PPV, which utilizes a multi-network architecture. Specifically, the proposed approach integrates an actor network that generates probabilistic policies. Additionally, two critic networks are incorporated to evaluate Q-values , thereby reducing the potential overestimation issues. The PFAM is integrated between hidden layers of the actor network to enhance feature representation. The current state $s_t$ serves as input to the network architecture, which subsequently produces action $a_t$. Upon action execution, the environment returns reward $r_t$, thereby forming a state transition. Subsequently, the VRC module handles this transition and saves it to the PER buffer for network training purposes.

\par Algorithm \ref{algo:SAC-PPV} presents the detailed procedure of the proposed SAC-PPV approach. Initially, the algorithm commences with parameter initialization, which includes actor and critic network weights, target network parameters, entropy coefficients, and PER buffer configurations. Subsequently, the system enters an iterative training phase that consists of multiple episodes. Each episode begins with environment and state reinitialization, followed by timestep-wise action sampling from the policy distribution and environmental execution. Upon action completion, the algorithm computes reward values and applies VRC module processing to these rewards. The resulting transitions are deposited into the PER buffer with assigned initial priority values. Once adequate transitions accumulate, the algorithm performs priority-based batch sampling from the buffer and calculates corresponding importance sampling weights. The critic networks undergo updates through TD error minimization, while the actor network optimization targets expected return and entropy maximization, with PFAM providing enhanced feature representations. Furthermore, target networks receive soft updates, and entropy coefficients are adjusted to maintain appropriate exploration levels. The replay buffer priorities are refreshed based on newly computed TD errors. This iterative process persists until all designated episodes conclude, enabling continuous policy refinement for optimal reinforcement learning performance.

\begin{algorithm} [tb]
  \caption{SAC-PPV}\label{algo:SAC-PPV}
    Initialize the parameters $\theta$ of the actor network, $\phi_1$ and $\phi_2$ of the critic networks, and $\alpha$ of the entropy coefficient;\\
    Initialize the target critic network parameters $\phi_{1,\text{target}} \leftarrow \phi_1$ and $\phi_{2,\text{target}} \leftarrow \phi_2$;\\
    Initialize the PER buffer $C$ with capacity $N$;\\

  \For{Episode $= 1, \ldots, N^{eps}$}
  {
    Reset the environment and initialize state $s_t$;\\
    \For{Time step $t = 1, \ldots, T$}
    {
        Obtain state $s_t$;\\
        Sample action $a_t \sim \pi_\theta(a_t|s_t)$ from the policy distribution;\\
        Execute action $a_t$ in the environment;\\
        Calculate $r_t$ according to Eq. ~\eqref{eq:reward};\\
        Process reward using the VRC module;\\
        Store the transition $(s_t, a_t, r_t, s_{t+1})$ in PER buffer $C$ with initial priority $p_i$;\\

        \If{Size(C) $\geq$ batch size}
        {  
            Retrieve a prioritized mini-batch of transitions based on $p_i$;\\  
            Update the critic by minimizing the loss as described in Eq.~\eqref{eq:sac_objective};\\  
            Update the actor with PFAM by maximizing the objective from Eq.~\eqref{eq:entropy_objective};\\
            Update the entropy coefficient $\alpha$ via gradient descent;\\  
            Soft update the target critic network:  
            \[
            \phi_{j,\text{target}} \leftarrow \tau \phi_j + (1 - \tau) \phi_{j,\text{target}} \text{ for } j=1,2;
            \]
            Update the priorities in the replay buffer based on new TD errors;\\
        }
    }
  }
\end{algorithm}

\subsection{Complexity Analysis}
\label{ssec:Complexity_Analysis}
\par This section analyzes the computational and space complexity characteristics of the SAC-PPV algorithm for both training and execution phases.
\par \textbf{Training Phase}: The computational complexity of SAC-PPV during training is expressed as $\mathcal{O}(N_A+N_{C_1}+ N_{C_2}+E \cdot T \cdot N_A+E \cdot T \cdot C_{PER} \cdot C_{env})+ E \cdot T / B \cdot (N_A+N_{C_1}+ N_{C_2})$. This comprises the following components \cite{Liang2025}:
\begin{itemize}
    \item \textbf{Network Initialization}: Parameter initialization for all networks yields computational complexity of $\mathcal{O}(N_A + N_{C_1}+ N_{C_2})$, where $N_A$ represents the parameter count in the actor network, while $N_{C_1}$ and $N_{C_2}$ denote the parameter counts in the two critic networks.
    
    \item \textbf{Action Sampling}: Action generation based on current state information requires computational complexity of $\mathcal{O}(E \cdot T \cdot N_A)$, where $E$ represents the total training episodes and $T$ denotes the timesteps per episode.
    
    \item \textbf{PER Buffer}: The complexity for this component is $\mathcal{O}(E \cdot T \cdot C_{PER} \cdot C_{env})$, where $C_{PER}$ denotes the computational overhead of the PER module and $C_{env}$ represents the environmental interaction complexity.
    
    \item \textbf{Network Update}: Parameter updates occur $K$ times when the replay buffer reaches batch size $B$. This yields computational complexity of $\mathcal{O}\left(E \cdot T / B \cdot K \cdot (N_A+N_{C_1}+ N_{C_2})\right)$.

\end{itemize}

\par For the training phase, space complexity is $\mathcal{O}\left(N_A+N_C+C_{buffer} \cdot (2d_s+d_a+2)\right)$, where $C_{buffer}$ denotes replay buffer capacity, $d_s$ and $d_a$ represent state and action space dimensions respectively. This includes both network parameter storage and memory allocation for transition tuples $\left(s_t, a_t, r_t, s_{t+1}\right)$ in the replay buffer.
\par \textbf{Execution Phase}: During deployment, the primary computational overhead stems from action generation by the actor network, yielding complexity $\mathcal{O}(N_A)$. The corresponding space complexity is also $\mathcal{O}(N_A)$.

\begin{table}
\caption{Summary of Parameters}
\label{tab:parameters}
\renewcommand{\arraystretch}{1.2}
\centering
\begin{tabular}{ll}
\toprule
\textbf{Parameter} & \textbf{Value} \\
\midrule
BLS Network size & $400 \mathrm{~m} \times 400 \mathrm{~m}$ \\
Initial position of UAV & $(100 \mathrm{~m},100 \mathrm{~m})$ \\
Number of BLSs & $20$ \\
Maximum flying distances of UAV $x_{\max}$ and $y_{\max}$ & $20 \mathrm{~m}$ \cite{Xie2024} \\
Path loss $\alpha$ & 2.6 \cite{Wan2024} \\
Noise Power $\sigma^2$ & $-110 \mathrm{~dbm}$ \cite{Pan2023} \\
Number of training episodes $N_e$ & 12000 \\
Learning rate & $3 \times 10^{-4}$ \\
Discounted factor $\gamma$ & 0.99  \\
Replay buffer size  & $10^6$  \\
Batch size & 256 \\
Soft update weight $\varepsilon$ & 0.005  \\
\bottomrule
\end{tabular}
\end{table}

%
%
\section{Simulation And Analyses}
\label{sec:Simulation And Analyses}
\par This section demonstrates the simulation results and corresponding analysis. Initially, we describe the simulation setting and baseline methods. Subsequently, we present the detailed findings from our experiments.
\subsection{Simulation Setups}
\par We consider a target area of size $400 \mathrm{~m} \times 400 \mathrm{~m}$, where a UAV serves 20 BLSs randomly distributed within the region. Moreover, the UAV operates within this area at a fixed altitude of 50 m. We set the bottom-left corner of the monitoring area as the origin, and the initial coordinates of the UAV are $(100 \mathrm{~m},100 \mathrm{~m})$. In each time slot, the data generation state of the sensors is refreshed, with the occurrence of new data governed by a Bernoulli distribution. When new data is generated, the volume of the data is modeled by a normal distribution. Additionally, more parameters related to the simulation environment are listed in Table~\ref{tab:parameters}.
\par For comparison, we utilize proximal policy optimization (PPO) \cite{Yuhui2025}, twin delayed deep deterministic policy gradient (TD3) \cite{Minghao2025}, truncated quantile critics (TQC) \cite{Kuznetsov2020}, and soft actor critic (SAC) \cite{Goudarzi2025} as benchmark methods.
\begin{itemize}
    \item \textit{PPO} \cite{Yuhui2025}: PPO is a trust region-based policy gradient algorithm that ensures training stability by limiting policy update steps through objective function clipping. The algorithm has simple implementation and low sensitivity to hyperparameters, with excellent performance in high-dimensional continuous action space tasks such as UAV control.
    \item \textit{TD3} \cite{Minghao2025}: TD3 is an improved version of DDPG that enhances training stability through the introduction of dual delayed update mechanisms, dual Q-networks, and target policy smoothing techniques. The algorithm performs excellently in continuous control tasks and can effectively handle high-dimensional states and action spaces in UAV trajectory planning.
    \item \textit{TQC} \cite{Kuznetsov2020}: TQC is an actor-critic algorithm based on quantile regression that reduces over-optimistic bias in Q-value estimates by truncating the highest quantiles. The algorithm demonstrates superior performance when processing complex reward structures and high-dimensional state spaces and is particularly suitable for decision-making tasks in UAV environments.
    \item \textit{SAC} \cite{Goudarzi2025}: SAC is an off-policy algorithm based on the maximum entropy reinforcement learning framework that balances exploration and exploitation by simultaneously optimizing stochastic policies and temperature parameters. SAC performs excellently in continuous control tasks with high sample efficiency and strong training stability and provides a reliable benchmark for UAV trajectory planning and power allocation.
\end{itemize}
\par Note that the server used for experiments is equipped with an Intel(R) Xeon(R) Gold 6330 CPU and an NVIDIA GeForce RTX 3060, with 125 GB of RAM. Additionally, the algorithm simulations above are conducted on Pytorch 2.4.1 and Python 3.8.20.

\subsection{Simulation Results}

\begin{figure}[!t]
  \centering
  \includegraphics[width=3.5in]{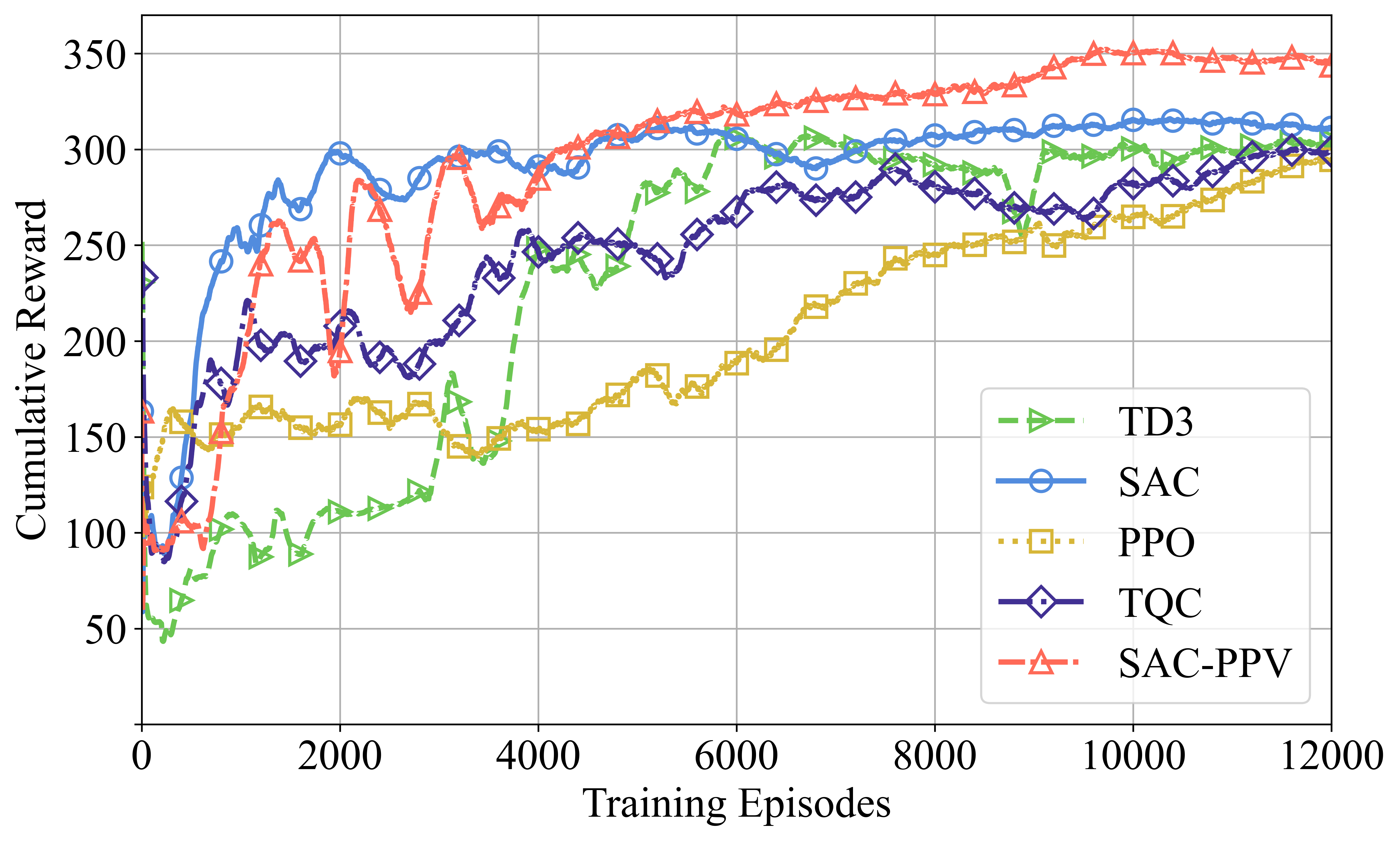}
  \caption{Cumulative rewards training curve.}
  \label{fig:reward}
\end{figure}

\begin{figure}[!t]
  \centering
  \includegraphics[width=3.5in]{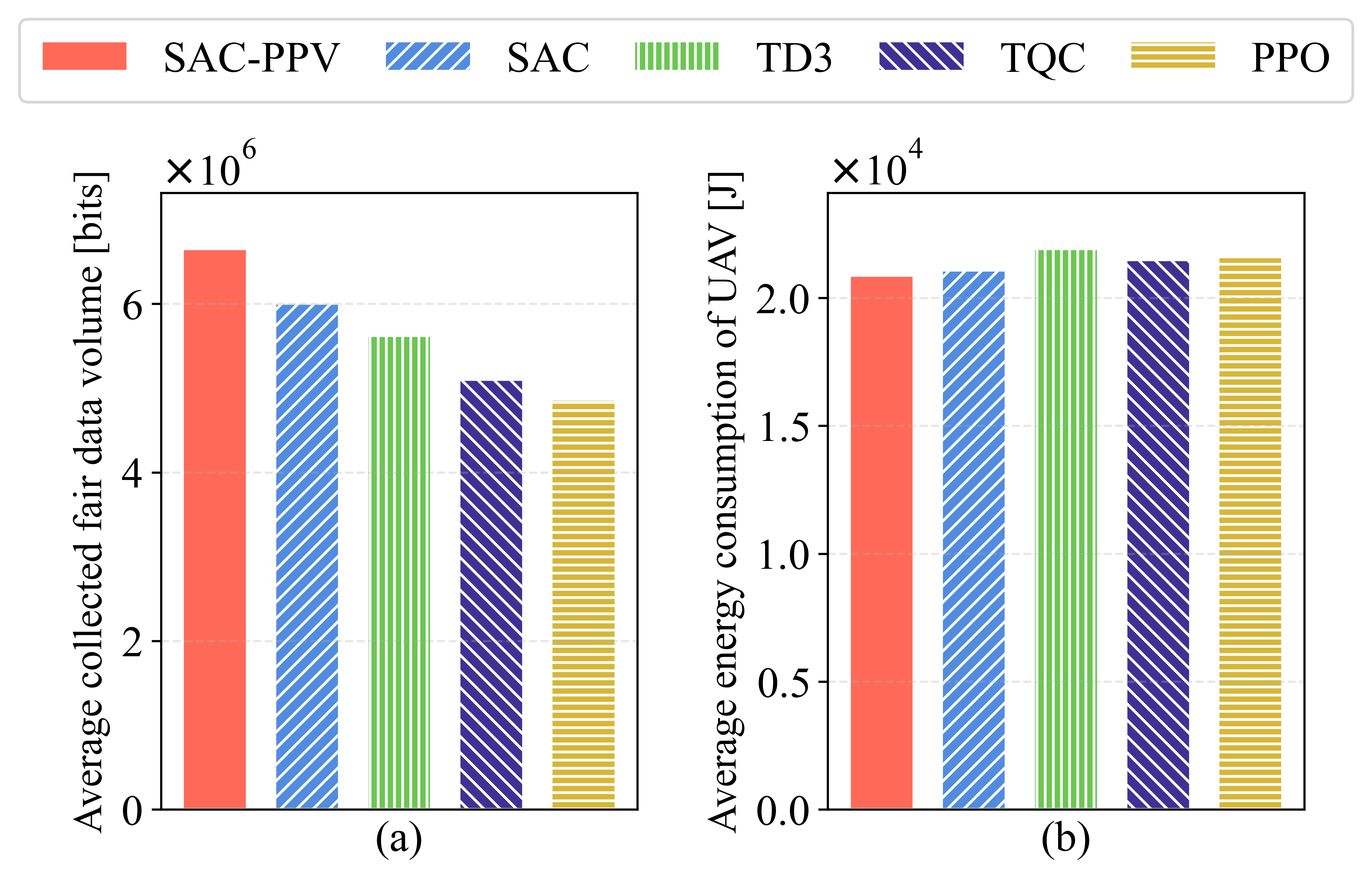}
  \caption{Training results when the network consists of 20 BLSs. (a) Average collected fair data volume [bits]. (b) Average energy consumption of UAV [J].}
  \label{fig:comparison_bar}
\end{figure}

\par \textit{1) Comparative Analysis of DRL Algorithm Performance:} Fig.~\ref{fig:reward} presents the reward convergence curves for all DRL algorithms. The results demonstrate that the proposed SAC-PPV algorithm provides higher rewards compared to other benchmark algorithms. This advantage primarily stems from the integration of the PFAM into the Actor network, which enables more effective feature extraction from the state space. Additionally, the incorporation of PER allows the algorithm to perform more efficient sampling. The implementation of VRC technology further contributes to faster and more stable convergence. Furthermore, the algorithms TD3 and TQC show competitive but lower performance compared to SAC throughout most of the training process. TD3 exhibits significant fluctuations during early episodes before eventually stabilizing, while TQC demonstrates more consistent but moderate performance. Moreover, it is worth noting that despite the initially slow convergence rate of PPO, this algorithm exhibits steady improvement throughout the entire training process without significant performance degradation. The robustness of PPO can be attributed to the conservative policy update mechanism, which prevents catastrophic performance drops but potentially results in reduced learning rates. Overall, the superior performance of SAC-PPV over the standard SAC algorithm indicates that the proposed modifications effectively enhance the learning capability and stability of the algorithm in this specific task.
\par To demonstrate the performance of the proposed algorithm in fair data collection, we plot the average fair data collection volume for the SAC-PPV algorithm and other benchmark algorithms in Fig.~\ref{fig:comparison_bar}. The experimental results demonstrate that the proposed SAC-PPV algorithm attains superior average fair data collection volume relative to baseline methods. This performance advantage stems from the fact that SAC-PPV incorporates three key enhancements. The PFAM effectively captures spatial relationships between UAVs and users, while the PER mechanism focuses training on more informative transitions. Furthermore, the VRC technique improves value estimation accuracy throughout the training process, resulting in superior overall performance. In contrast, the standard SAC algorithm lacks these optimization components, while TD3 suffers from instability during training despite its twin networks. Additionally, TQC demonstrates moderate performance due to its truncated quantile approximation, and PPO exhibits the lowest performance because its on-policy nature limits sample efficiency in this complex environment.

\begin{figure*}
  \centering
  \includegraphics[width=1\textwidth]{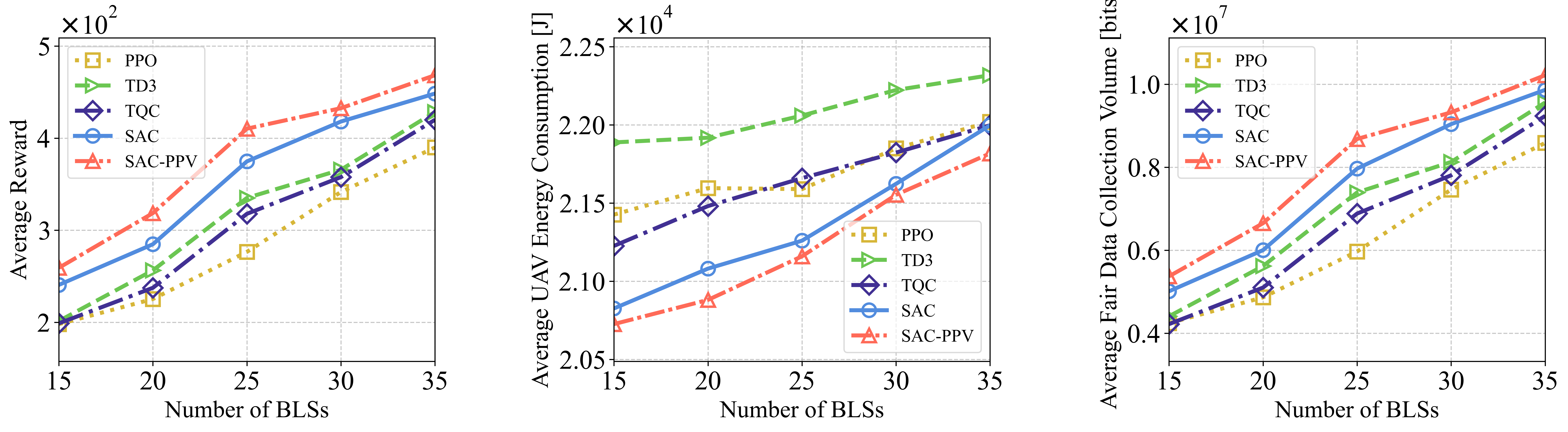}
  \caption{The effect of the number of BLSs on (a) average reward (b) average UAV energy consumption [J] (c) average fair data collection volume [bits].}
  \label{fig:sensor_number_analysis}
\end{figure*}

\begin{figure}[!t]
  \centering
  \includegraphics[width=3.5in]{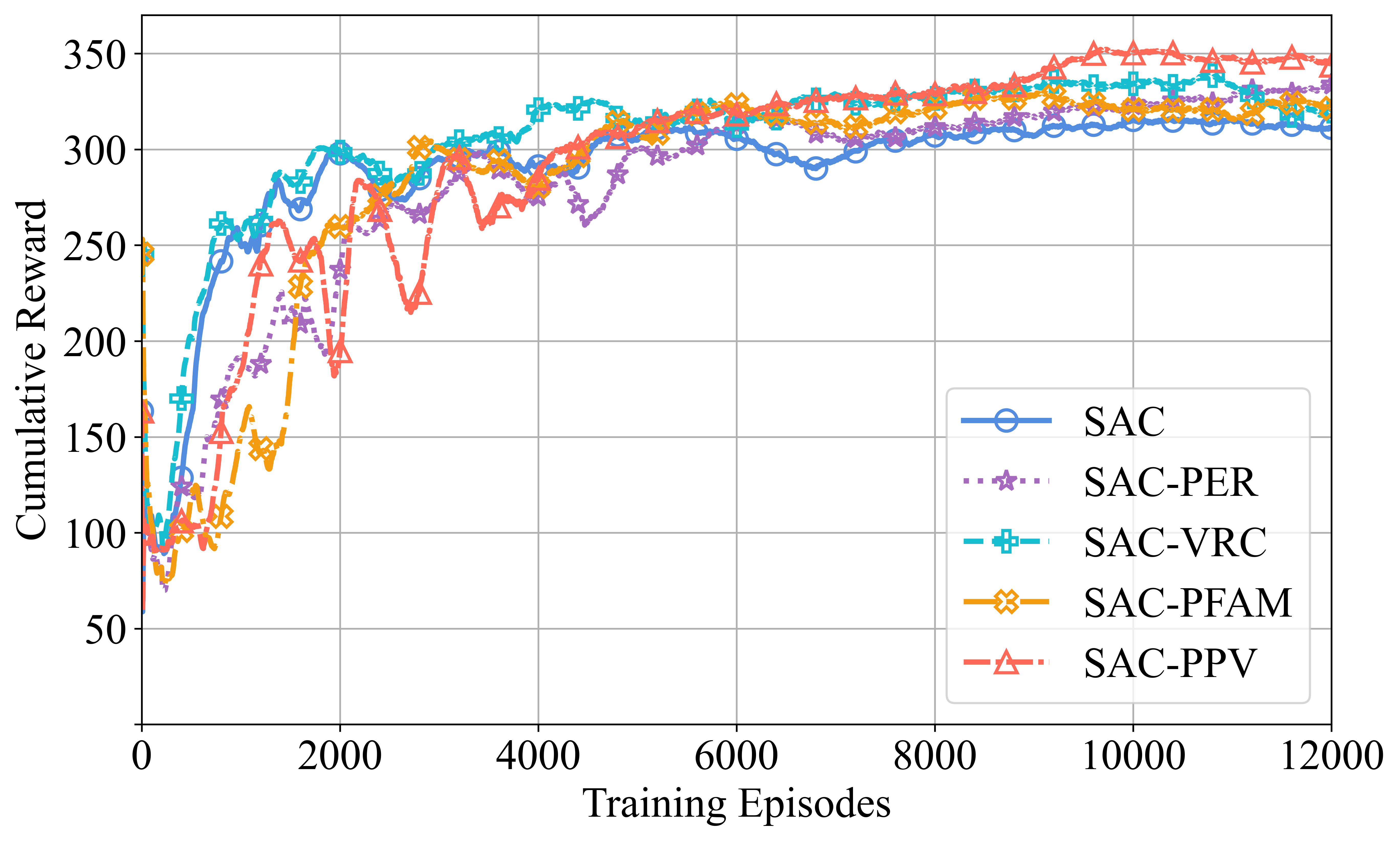}
  \caption{Ablation simulation results when the network consists of 20 BLSs. Comparison between SAC-PPV and standard SAC, SAC with PFAM, SAC with PER, and SAC with VRC.}
  \label{fig:ablation_plot}
\end{figure}

\par Moreover, we assess the performance of these algorithms in solving the proposed problems via UAV power consumption, as illustrated in Fig.~\ref{fig:comparison_bar}. As can be seen, SAC-PPV achieves the lowest UAV energy consumption by learning more effective energy management strategies. The reason for this may be that the PFAM enables more efficient path planning while the PER mechanism helps the algorithm learn energy-conserving maneuvers from critical experiences. Additionally, the VRC technique allows UAVs to coordinate movements that reduce unnecessary energy expenditure.
\par \textit{2) Ablation Simulation Results:} Fig.~\ref{fig:ablation_plot} illustrates our ablation experiment results comparing the baseline SAC algorithm with its enhanced variants. The integration of PFAM with SAC demonstrates significant improvement in both convergence speed and final rewards because PFAM enables more efficient feature extraction through its adaptive attention mechanism. This result confirms the effectiveness of attention-based feature processing in reinforcement learning tasks. Furthermore, SAC combined with PER exhibits enhanced learning stability and reduces variance during training due to the prioritized sampling approach that focuses on more informative experiences. This result validates the importance of experience replay optimization in complex environments. Additionally, SAC with VRC shows remarkable performance gains in terms of reward accumulation because the value prediction component provides more accurate future state estimations. This result substantiates the advantage of incorporating predictive modeling within the learning framework. When all three enhancement features are jointly utilized, the algorithm achieves the highest overall performance with faster convergence and superior final rewards, demonstrating the complementary nature of these components and their synergistic effect on reinforcement learning performance in challenging control tasks.
\par \textit{3) Impact of BLS Quantity:} The quantity of BLS significantly affects system performance and resource allocation efficiency. Therefore, in Fig.~\ref{fig:sensor_number_analysis}, we illustrate the variations in reward, fair data collection volume, and UAV energy consumption relative to the number of sensors $N_{SN}$. We observe that fair data collection volume demonstrates an upward trend with increasing $N_{SN}$. This occurs because more sensors provide additional data collection opportunities across the environment, enhancing overall system coverage. However, the increased number of BLS requires more data collection and charging operations, which consequently increases UAV energy consumption. Notably, our proposed SAC-PPV algorithm outperforms other algorithms in both reward metrics and fair data collection volume. These improvements result from the integration of PFAM, PER, and VRC, which enable agents to make more balanced decisions between fairness and energy efficiency, thereby optimizing overall system performance. Interestingly, the UAV energy consumption of SAC-PPV exceeds that of the PPO algorithm. This happens because PPO exhibits more conservative behavior and lower velocity profiles, resulting in reduced charging of BLS nodes and data collection from BLS nodes, thus achieving minimal energy consumption. Nevertheless, this conservative strategy lacks proper planning for fair resource allocation, leading to poor performance in fair data collection. Overall, compared to other algorithms, the SAC-PPV algorithm demonstrates robust adaptability to increasing numbers of BLS nodes.

\begin{figure}[!t]
  \centering
  \includegraphics[width=3.5in]{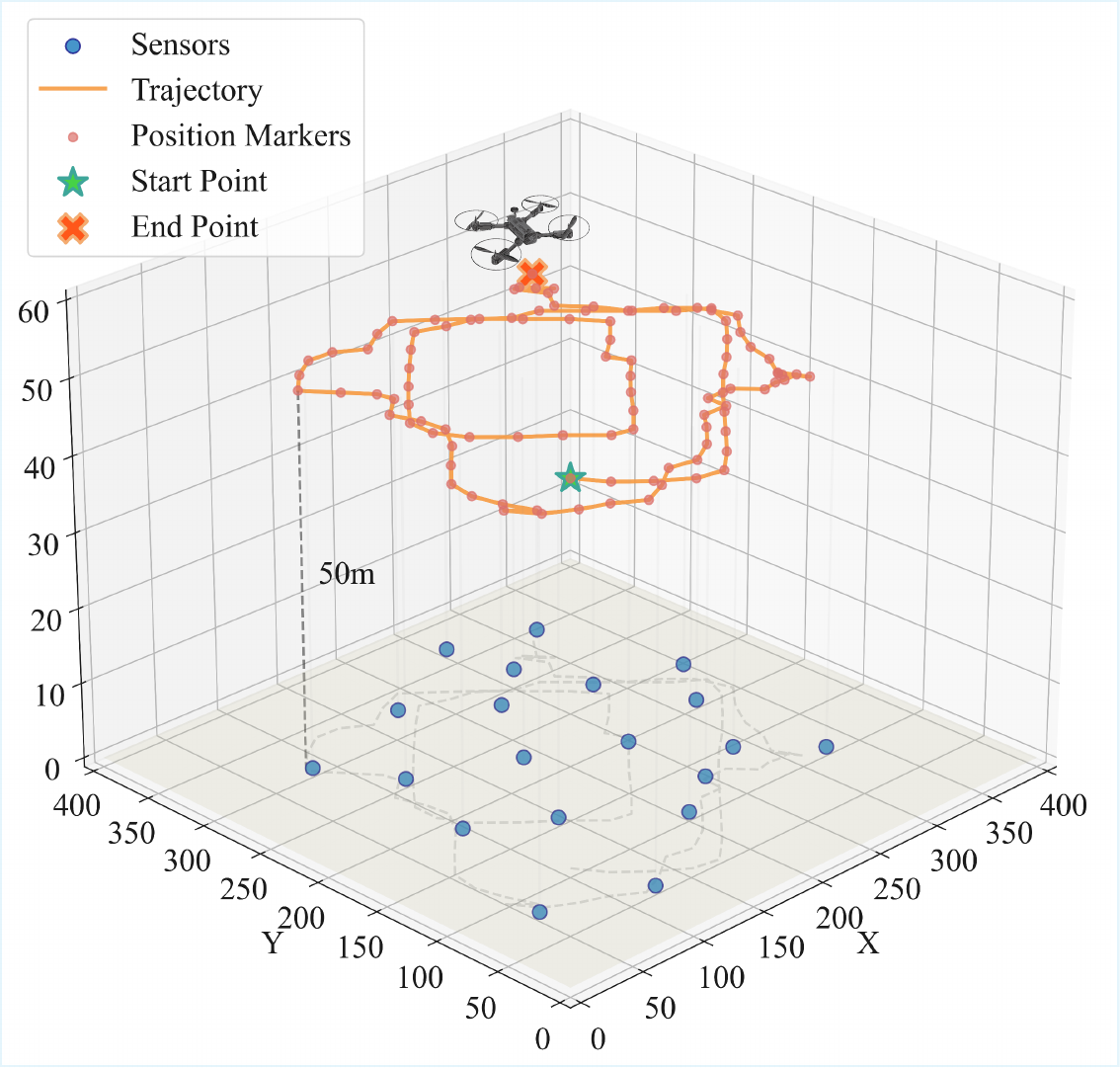}
  \caption{The UAV trajectory trained by SAC-PPV when the network consists of 20 BLSs.}
  \label{fig:trajectory}
\end{figure}

\par \textit{4) UAV Trajectory Analysis:} From Fig.~\ref{fig:trajectory}, we can observe that the UAV trajectory effectively covers the entire monitoring area with multiple traversal loops. This multi-pass strategy occurs because BLS nodes generate new data according to a Bernoulli distribution, necessitating the UAV to return to previously visited areas for additional data collection. Additionally, the figure demonstrates that the UAV exhibits more concentrated movements in regions with higher sensor density, which indicates the algorithm can demonstrate excellent adaptive behavior in response to sensor distribution. Moreover, the UAV flies to BLS locations, which demonstrates a strong emphasis on fairness in data collection rather than merely minimizing energy consumption. Overall, the SAC-PPV algorithm controls the UAV to achieve a trajectory that balances fair data collection volume and UAV energy consumption, while adaptively responding to the dynamic nature of sensor data generation across the operational environment.

%
%
\section{Conclusion}
\label{sec:Conclusion}

\par This paper has investigated a UAV-assisted data collection and WPT system for BLS networks in remote areas. We have proposed a framework wherein a UAV equipped with WPT modules first charges the BLS nodes, after which these nodes utilize the harvested energy to transmit their sensed data to the UAV through OFDMA. Within this system, we have formulated a multi-objective optimization problem that aims to maximize fair data collection volume while minimizing UAV energy consumption through joint optimization of transmit power allocation and flight trajectory planning. Subsequently, the problem has been reformulated as a continuous-space MDP to enhance computational efficiency and practical implementation. Then, we have proposed the SAC-PPV algorithm that incorporates PFAM for adaptive feature extraction, PER for efficient learning from critical samples, and VRC to stabilize training in dynamic WPT environments. Simulation results have demonstrated that our proposed SAC-PPV algorithm has outperformed various benchmarks by achieving maximum fair data collection volume while maintaining low UAV energy consumption throughout different operational scenarios. This approach has provided an effective solution for joint optimization of UAV transmit power and trajectory in WPT-enabled BLS networks, which has proved particularly beneficial for remote IoT applications with stringent energy constraints and limited infrastructure access.

\bibliographystyle{IEEEtran}
\bibliography{ref}

\begin{IEEEbiography}[{\includegraphics[width=1in,height=1.25in,clip,keepaspectratio]{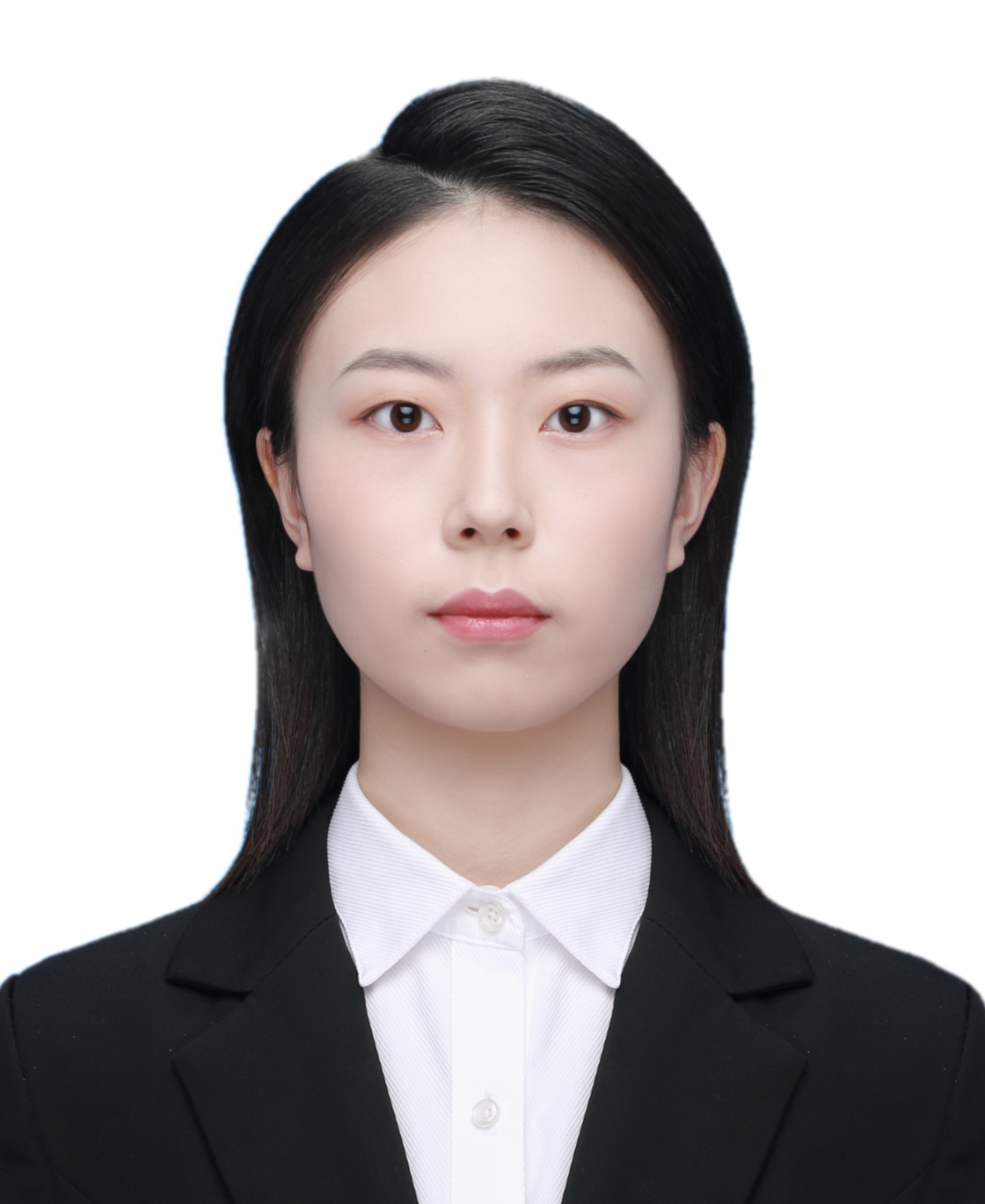}}]{Wen Zhang}
received the BS degree in computer science and technology from Jilin University, Changchun, China, in 2024. She is currently working toward the Ph.D. degree in computer technology with Jilin University, Changchun, China. Her research interests are UAV networks and optimization.
\end{IEEEbiography}

\begin{IEEEbiography}[{\includegraphics[width=1in,height=1.25in,clip,keepaspectratio]{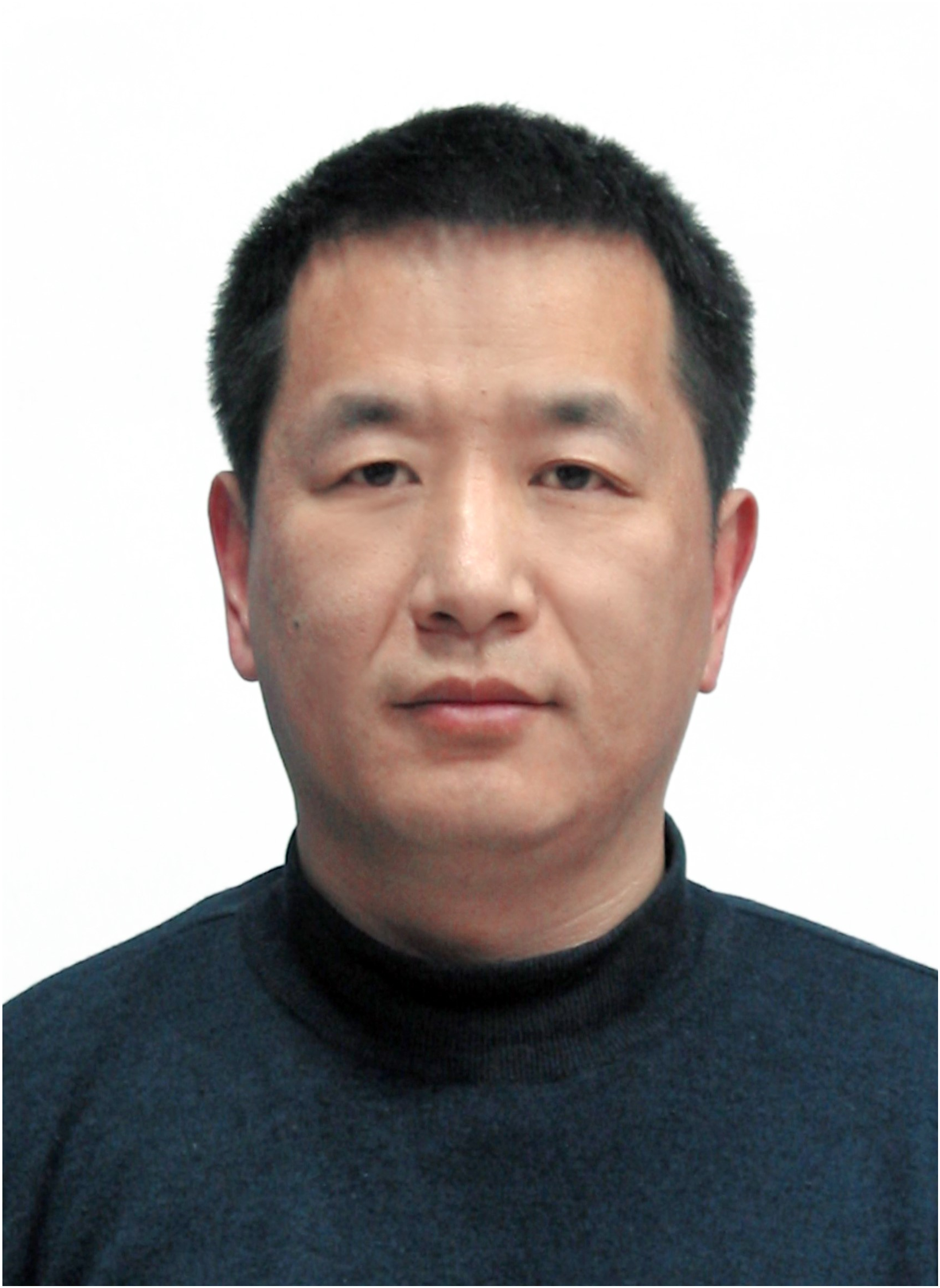}}]{Aimin Wang}
received Ph.D. degree in Communication and Information System from Jilin University. He is currently a professor at Jilin University. His research interests are wireless sensor networks and QoS for multimedia transmission.
\end{IEEEbiography}

\begin{IEEEbiography}[{\includegraphics[width=1in,height=1.25in,clip,keepaspectratio]{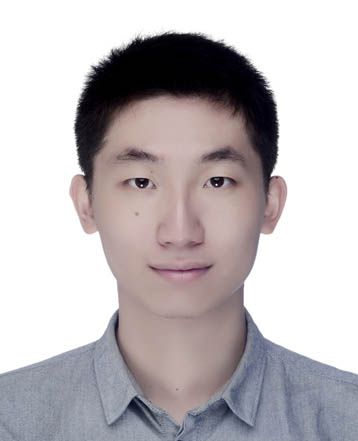}}]{Jiahui Li}
  (Member, IEEE) received his B.S. in Software Engineering, and M.S. and Ph.D. in Computer Science and Technology from Jilin University, Changchun, China, in 2018, 2021, and 2024, respectively. He was a visiting Ph.D. student at the Singapore University of Technology and Design (SUTD). He currently serves as an assistant researcher in the College of Computer Science and Technology at Jilin University. His current research focuses on integrated air-ground networks, UAV networks, wireless energy transfer, and optimization.
\end{IEEEbiography}

\begin{IEEEbiography}[{\includegraphics[width=1in,height=1.25in,clip,keepaspectratio]{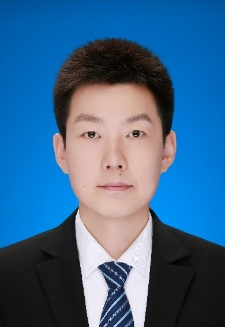}}]{Geng Sun} 
(Senior Member, IEEE) received the B.S. degree in communication engineering from Dalian Polytechnic University, in 2007, and the Ph.D. degree in computer science and technology from Jilin University, in 2018, respectively. He was a Visiting Researcher with the School of Electrical and Computer Engineering, Georgia Institute of Technology, USA. He is a Professor in College of Computer Science and Technology at Jilin University, and his research interests include wireless networks, UAV communications, collaborative beamforming and optimizations.
\end{IEEEbiography}

\begin{IEEEbiography}
[{\includegraphics[width=1in,height=1.25in,clip,keepaspectratio]{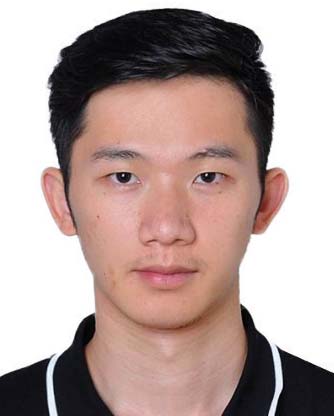}}]{Jiacheng Wang} 
received the Ph.D. degree from the School of Communication and Information Engineering, Chongqing University of Posts and Telecommunications, Chongqing, China. He is currently a Research Associate in computer science and engineering with Nanyang Technological University, Singapore. His research interests include wireless sensing, semantic communications, and metaverse.
\end{IEEEbiography}

\begin{IEEEbiography}
[{\includegraphics[width=1in,height=1.25in,clip,keepaspectratio]{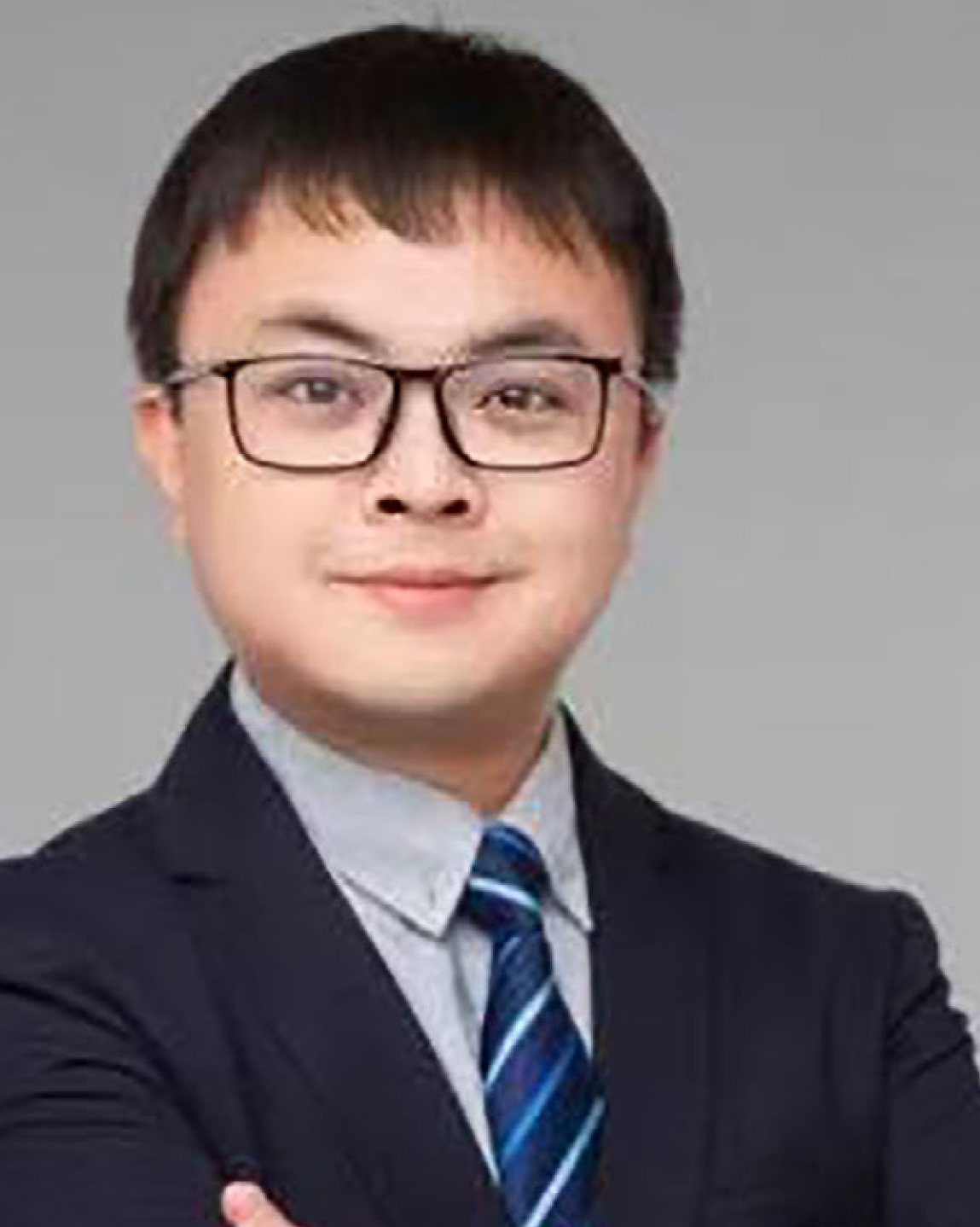}}]{Weijie Yuan} 
(Senior Member, IEEE) received the Ph.D. degree from the University of Technology Sydney, Australia, in 2019. In 2016, he was a Visiting Ph.D. Student with the Institute of Telecommunications, Vienna University of Technology, Austria. He was a Research Assistant with The University of Sydney, a Visiting Associate Fellow with the University of Wollongong, and a Visiting Fellow with the University of Southampton, from 2017 to 2019. From 2019 to 2021, he was a Research Associate with the University of New South Wales. He is currently an Assistant Professor with Southern University of Science and Technology. He currently serves as an Associate Editor for IEEE TRANSACTIONS ON WIRELESS COMMUNICATIONS, IEEE Communications Standards Magazine, IEEE TRANSACTIONS ON GREEN COMMUNICATIONS AND NETWORKING, IEEE COMMUNICATIONS LETTERS, IEEE OPEN JOURNAL OF COMMUNICATIONS SOCIETY, and EURASIP Journal on Advances in Signal Processing. He is the Lead Editor for the series on ISAC in IEEE Communications Magazine. He was an Organizer/the Chair of several workshops and special sessions on OTFS and ISAC in flagship IEEE and ACM conferences, including IEEE ICC, IEEE/CIC ICCC, IEEE SPAWC, IEEE VTC, IEEE WCNC, IEEE ICASSP, and ACM MobiCom. He is the Founding Chair of the IEEE ComSoc Special Interest Group on OTFS (OTFS-SIG). He has been listed in the World’s Top 2\% Scientists by Stanford University for citation impact since 2021. He was a recipient of the Best Editor Award from IEEE COMMUNICA-TIONS LETTERS, the Exemplary Reviewer from IEEE TRANSACTIONS ON COMMUNICATIONS/IEEE WIRELESS COMMUNICATIONS LETTERS, and the Best Paper Award from IEEE ICC 2023, IEEE/CIC ICCC 2023, and IEEE GlobeCom 2024.
\end{IEEEbiography}

\begin{IEEEbiography}
[{\includegraphics[width=1in,height=1.25in,clip,keepaspectratio]{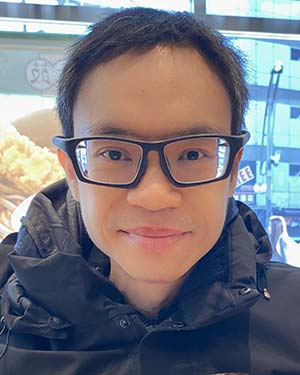}}]{Dusit Niyato} 
(Fellow, IEEE) received the B.Eng. degree from the King Mongkuts Institute of Technology Ladkrabang (KMITL), Thailand, in 1999, and the Ph.D. degree in electrical and computer engineering from the University of Manitoba, Canada, in 2008. He is currently a Professor with the School of Computer Science and Engineering, Nanyang Technological University, Singapore. His research interests include the Internet of Things (IoT), machine learning, and incentive mechanism design.
\end{IEEEbiography}

\end{document}